\documentclass{article}
\pdfoutput=1 

\usepackage{ChiMacro}
\usepackage{arxiv}
\usepackage{csquotes}
\title{Vehicle Dynamics Control for Simultaneous Optimization of Tire Emissions and Performance in EVs}

\author{
  Chi Bach Pham 
    \qquad Homayoun Hamedmoghadam\qquad Robert Shorten\thanks{Corresponding author \newline Email address: r.shorten@imperial.ac.uk}\\
  Dyson School of Design Engineering\\
  Imperial College London\\
  London, SW7 2AZ \\
}

\begin{document}
\maketitle
\thispagestyle{fancy}
\begin{abstract}
In recent years, Electric Vehicles (EVs) have seen widespread public adoption. While EVs produce zero tailpipe emissions, they contribute to an increase in another type of vehicular emission: tire emissions. Battery-operated EVs are generally heavier than their combustion-engine counterparts and require greater acceleration forces, which their high-torque electric motors provide. This combination of increased weight and traction forces leads to higher tire emissions, which possess various adverse health and environmental effects. Here, we propose a control solution with promising results in mitigating tire wear in all-wheel-drive EVs. The idea is to utilize different tire profiles on each drive axis: a low-wear, low-traction axis and a high-wear, high-traction axis. Derived from detailed mathematical analyses, we propose a simple control scheme to counteract the performance difference from using the low-traction tires. The proposed control mechanism then distributes torque optimally between the two axes, maximizing usage from the low-wear axis and simultaneously maintaining stability and performance by leveraging high-traction tires. Through detailed numerical simulations, we demonstrate that the developed model significantly reduces tire emissions and maintains vehicle drivability and performance.

\end{abstract}

\section{Introduction}
The excessive emission of microplastic into the environment and its detrimental effects have been frequent topics of discussion for the last decades. A type of microplastics that has gained much traction in recent years is rubber particles emitted from vehicle tires, which is an unavoidable cost for safe operation of vehicles  ~\cite{luo2021environmental,sommer2018tire,tamis2021environmental,baensch2020tyre,wagner2018tire,timmers2016non}. As a matter of fact, tire wear is one of the largest sources of microplastics emissions~\cite{wagner2018tire}, with approximately 1.327 million metric tons of Tire Wear Particles (TWP) released into the environment in Europe in 2014, and 1.120 million metric tons of TWP generated from U.S. roads only in 2010~\cite{wagner2018tire}. TWPs are first deposited into the environment, and after settling, may be deposited again and again into the air through 
turbulence from traffic and wind~\cite{JARLSKOG2021145503,GNECCO200560}. The wear particles may continue to travel into the water resources through sewers, storm drains, and nearby natural bodies of water~\cite{kole2017wear,sommer2018tire}. The adverse health and environmental effects of the increase in the concentration of microplastics in the environment cannot be understated~\cite{wik2009occurrence,saha2024effect,feng2024long,dalmau2020tire,wang2024review}. 

The concern regarding tire particles has become even more prominent due to the rapid adoption of electric vehicles (EVs) in recent years, growing five times into 26 million EVs during a four-year period from 2018 to 2022~\cite{IEA}. Although EVs do not produce harmful tailpipe emissions like their combustion-engine counterparts, their relatively higher weight and acceleration forces result in increased tire emissions~\cite{sommer2018tire, kaul2009traffic, kole2017wear, obereigner2020low}. The tire emission problem is expected to be only exacerbated by the progressively increasing weight of the new EV batteries~\cite{wieberneit2024optimal} and the rapidly growing EV presence on urban roads as the demand for a greener vehicle fleet continues to grow in many countries~\cite{IEA,EEA}. 

To date, various approaches have been proposed to address the problem of tire emissions from different angles ~\cite{gehrke2023mitigation}. 
Some authors have focused on optimizing the tire itself through developing new tire materials with improved characteristics, such as replacing the binding material of tires with silica~\cite{OECD,pratico2020particulate} instead of the current option of carbon black. Silica as a binding material possesses better abrasion resistance properties while minimizing energy loss and maintaining wet grip capability~\cite{ten2002silica,pratico2020particulate}. Others have suggested introducing policies to limit the use of winter tires~\cite{furuseth2020reducing, pratico2020particulate} or studded tires~\cite{furuseth2020reducing}, which have worse emission properties~\cite{sundt2016primary,gustafsson2008properties}. Reducing the vehicle weight is another approach, due to the linear relationship between the amount of wear particle generated and the vehicle weight~\cite{kole2017wear}. This has been suggested by various author~\cite{furuseth2020reducing,andersson2020microplastics,boulter2006road}. However, the development of cars with reduced weight is a complex challenge, including engineering necessary light-weight material with adequate strength and economic viability.

Another angle to tackle the TWP emission problem is through optimizing driver behavior and vehicle dynamics. An important factor influencing the amount of TWP generation is vehicle speed~\cite{johannesson2022potential, foitzik2018investigation},
Hence, several approaches have aimed at reducing the operating speeds, such as road designs that encourage lower speeds or driving assistance systems that monitor individual trips and provide positive feedback to help drivers improve their driving behavior~\cite{andersson2020microplastics, furuseth2020reducing}.  
Other works have proposed more binding solutions that do not suffer from drivers noncompliance as those suggested above. Examples include imposing speed limits and emission thresholds through regulatory frameworks~\cite{timmers2016non} and incorporating technologies that physically limit vehicle speed ~\cite{boulter2006road}. However, speed and acceleration performance are some the main points justifying the price of a vehicle, so there is probably little interest in adopting these limiting measures.

Arguably, the most recent approach is the development of algorithms that can distribute the torque generated from the motor to reduce the peak force experienced on the tires of the vehicle. In this line of work, Papaioannou et al.~\cite{papaioannou2024reducing,papaioannou2022optimal} developed a trajectory optimization algorithm that minimizes tire wear for automated articulated vehicles. Their findings suggest that a small increase in journey time significantly decreases tire wear. For non-automated vehicles, Gao et al.~\cite{gao2022torque} proposed a distribution algorithm to reduce tire wear by efficiently allocating driving torques in an all-wheel-drive BEV, taking into account engine efficiency and energy losses caused by tire slip. Obereigner et al.~\cite{obereigner2020low} also developed a torque distribution algorithm that not only reduces tire wear but also considers improving the traffic capacity of roads.

A drawback of many studies involving optimizing the vehicle is that they fundamentally alter the performance of the vehicle, making them unlikely to be embraced by vehicle users and manufacturers. 
In this paper, we propose an approach that leverages the benefits of two tire types with opposing characteristics (both commonly available on the market): one with high traction and high emissions (referred to as the soft tire) and another with low traction and low emissions (referred to as the hard tire). The idea hinges on a control mechanism that unevenly divides the total torque between the two axes, which optimally distributes forces on the wheels to reduce tire emission. An advantage of the control mechanism is its ability to correct for the performance difference that using dual tire types makes, and as a result, the low-emitting drive will feel almost exactly as if the car was driving on high-performing high-emission tires on all wheels.

The remainder of the paper is organized as follows. An overview of tire wear and the trade-off in optimizing traction versus tire wear is provided in~\cref{sec:pre}, the proposed model is detailed in~\cref{sec:method}, and the simulation setup and results are presented in~\cref{sec:res}. Finally, our findings are discussed and concluded in~\cref{sec:con}.
\section{Preliminaries}
\label{sec:pre}

In this section, we provide details on the vehicle model, providing the physics governing the system comprising the vehicle and its tires. The main challenge in performance control stems from the traction difference between the front and rear axes, which is the idea that allows for tire-emission reduction. Here, we introduce the model that formulates the vehicle's movement on a 2-dimensional plane. The presented model is a commonly used framework and will be used as the basis of the control system proposed in~\cref{sec:method}.

The vehicle lateral control, in the present work, is built upon the mathematical vehicle model known as the \enquote{bicycle model}, where each wheel axis is represented by a single wheel (and the front wheel can be steered).
The model uses the following constants: Yaw inertia $I_z$ (kg·m$^2$), drag coefficient $C_d$ (kg/m), wheel radius $r_e$ (m), wheel inertia $J$ (kgm$^2$), and the distances from the center of gravity (CoG) to the front and rear axes denoted respectively by $l_f$ (m) and $l_r$ (m). The list of all constants and our notations are display in~\cref{tab:const}.

\begin{figure}[hb]
    \centering
    \captionsetup{width=1\textwidth}
    \includegraphics[width=0.5\textwidth]{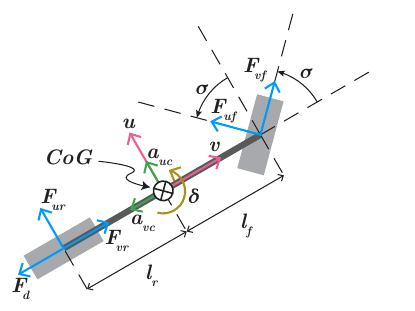}
    \captionsetup{justification=centering}
    \caption{ \textbf{Bicycle vehicle model.} Vehicle model with two-wheel axis, with steerable front wheel. The distance between the vehicle's center of gravity(CoG) to its front and rear axis are denoted $l_f$ and $l_r$, respectively. The vehicle is under five forces (colored in blue): Longitudinal and lateral forces from the front wheel $F_{v,f},\,F_{u,f}$, longitudinal and lateral forces from the rear wheel $F_{v,r},\,F_{u,r}$, as well as the air drag forces $F_d = C_dv^2$. The vehicle's longitudinal and lateral velocity are denoted $v$ and $u$, colored in red. The vehicle also has an angular velocity around its vertical $z$-axis, denoted $\delta$ and colored in yellow. This angular velocity along with the lateral and longitudinal velocities also created centripetal acceleration $a_{v,c}$ and $a_{u,c}$, colored in green. 
    }
    \label{fig:2axismodel}
    \vspace{-\belowdisplayskip}
\end{figure}

The vehicle's steering angle relative to the longitudinal velocity vector is denoted by $\sigma$ (rad).
The forces acting on the vehicle are: Drag force $F_d = C_dv^2$ (N), 
lateral front and rear tire force $F_{u,f}, F_{u,r}$ (N), 
longitudinal front and rear tire force $F_{v,f},\,F_{v,r}$ (N).
The longitudinal velocity of the vehicle is denoted by $v$~(m$\cdot$s$^{-1}$), the lateral velocity is denoted by $u$~(m$\cdot$s$^{-1}$), and the angular velocity around the $z$-axis (also commonly refers to as the yaw rate) is denoted by $\delta$ (rad$\cdot$s$^{-1}$).
Since the vehicle is also under spinning moment along its $z$ axis, it is also affected by centripetal acceleration from longitudinal velocity and lateral velocity, which are represented by $a_{v,c} = v\delta$~(m$\cdot$s$^{-2}$) and $a_{u,c} = u\delta$~(m$\cdot$s$^{-2}$). The sketch in~\cref{fig:2axismodel} visualizes the components of the bicycle model, with all its parameters as an aid to facilitate following the formulations coming up next.
\begin{table}
 \caption{Summary of constants used in the \enquote{Bicycle Model} of vehicle dynamics.}
  \centering
  \begin{tabular}{lll}
    \toprule               
    Variable & Description & Unit \\
    \midrule
    $m$ & Vehicle weight & kg\\
    $g$ & Gravitational acceleration &ms$^{-2}$\\
    $I_z$ & Yaw inertia  & kgm$^2$    \\
    $C_d$     & Drag coefficient & kgm$^{-1}$      \\
    $r_e$     & Wheel radius       & m  \\
    $J$         & Wheel inertia  & kgm$^2$\\
    $l_f$      &Distance from front axis to center of gravity & m\\
    $l_r$&Distance from rear axis to center of gravity & m\\
    \bottomrule
  \end{tabular}
  \label{tab:const}
\end{table}%

Applying Newton's second law on vehicle longitudinal forces, we get the following expression for the vehicle longitudinal acceleration:
\begin{align}
    \frac{dv}{dt} = \frac{1}{m}\left(F_{v,r} + F_{v,f}\cos(\sigma) - F_{u,f}\sin(\sigma) - C_dv^2\right) + u\delta.\label{eq:dvdt}
\end{align}

Applying Newton's second law, this time on vehicle lateral forces, we get the following expression for the vehicle lateral acceleration:
\begin{align}
\frac{du}{dt}  = \frac{1}{m}\left(F_{u,r} + F_{v,f}\sin(\sigma) + F_{u,f}\cos(\sigma)\right) - v\delta.\label{eq:dudt}
\end{align}

Applying Newton's second law for rotation, we get the the vehicle yaw acceleration expressed as:
\begin{align}
    \frac{d\delta}{dt} = \frac{1}{I_z}\brac{-F_{u,r}l_r + l_f\brac{F_{v,f}\sin(\sigma) + F_{u,f}\cos(\sigma)}}.\label{eq:dpsidt2}
\end{align}
The longitudinal tire forces $F_{v,r}$ and $F_{v,f}$ used in~\cref{eq:dvdt,eq:dudt,eq:dpsidt2} is express as a function of longitudinal slip, which is the ratio between the difference between the speed of the vehicle and speed of the wheel over the speed of the vehicle, i.e., 
\begin{align}
    \lambda_k &= \frac{r_e\omega_k-v}{v}\label{eq:lambda}\\
    \frac{d\omega_k}{dt} &= \frac{T_k-F_{v,k}r_e}{J_k}\label{eq:domegadt},
\end{align}
where $\lambda_k, T_k, \omega_k, F_{v,k}$ are respectively slip ratio, engine torque output, wheel angular velocity, wheel inertia, and longitudinal tire force. The parameter $r_e$ indicates the wheel radius, the parameter $J_k$ indicates the wheel inertia, and the index $k\in\{f,r\}$ specifies the front ($k=f$) or rear ($k=r$) wheel.
The lateral tire forces $F_{u,r}$ and $F_{u,f}$ used in~\cref{eq:dvdt,eq:dudt,eq:dpsidt2} are expressed as a function of slip angle $\alpha$, which is defined as the difference in the heading angle of the tire and the vehicle. The slip angle for the front and rear wheels are formulated as
\begin{align}
    \alpha_f &= \sigma-\arctan(\frac{u+l_f\delta}{v}) \approx \sigma-\frac{u+l_f\delta}{v}\label{eq:alphaf}\\
    \alpha_r &= -\arctan(\frac{u-l_r\delta}{v})\approx -\frac{u-l_r\delta}{v}.\label{eq:alphar}
\end{align}
The lateral and longitudinal tire forces are expressed as functions of slip angle and slip ratio as follow:
\begin{alignat}{2}
    &F_{v,f} = \frac{mgl_r}{l_f+l_r}\cM_{v}(\lambda_f),\quad&&
    F_{v,r} = \frac{mgl_f}{l_f+l_r}\cM_{v}(\lambda_r)\\
    \label{eq:Fu}
    &F_{u,f} = \frac{mgl_r}{l_f+l_r}\cM_{u}(\alpha_f),\quad&&
    F_{u,r} = \frac{mgl_f}{l_f+l_r}\cM_{u}(\alpha_r),
\end{alignat}
where $\cM_{v}(.)$ and $\cM_{u}(.)$ are tire models, which respectively return the longitudinal and lateral tire friction coefficient.
Tire models are generated from tire and road interactions by using a standard empirical equation, known as \enquote{Magic Formula} by Pacejka~\cite{pacejka2005tire}: 
\begin{align}
    \mathcal{M}_{u}(\alpha) &= D_{u}\sin(C_{u}\tan^{-1}(B_{u}\alpha-E_{u}(B_{u}\alpha-\tan^{-1}(B_{u}\alpha)))),\nonumber\\
    \mathcal{M}_{v}(\lambda) &= D_{v}\sin(C_{v}\tan^{-1}(B_{v}\lambda-E_{v}(B_{v}\lambda-\tan^{-1}(B_{v}\lambda)))),
    \label{eq:magic}
\end{align}
which describes the lateral and longitudinal friction coefficient as a function of the appropriate slip value. In \cref{eq:magic}, $B_u,B_v,C_v,C_v,D_u,D_v,E_u,E_v$ are fitting parameters. For each tire, the fitting parameters are found by fitting the Magic Formula~(\cref{eq:magic}) to real measurement data. In this work, we obtain $B_u,B_v,C_v,C_v,D_u,D_v,E_u,E_v$ from fitting measurement data from a standard MATLAB tire model as will be explained in~\cref{sec:res}.
Among these parameters, $D_u$ and $D_v$ are the most important as they determine the maximum possible friction coefficient with $\max_{\alpha} \cM_u(\alpha) = D_u$ and $\max_{\lambda} \cM_v(\lambda) = D_{v}$. 
\begin{figure}[b]
    \centering
    \captionsetup{width=0.6\textwidth}
    \includegraphics[width=0.4\linewidth]{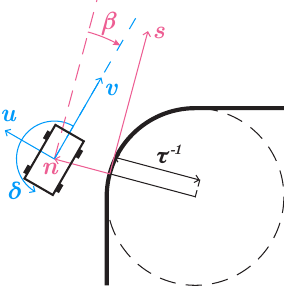}
    \captionsetup{justification=centering}
    \caption{\textbf{Curvilinear coordinate system.}
    The figure shows the vehicle frame coordinates (in blue) and curvilinear frame coordinates (in red) in relation to one another.}
    \label{fig:curvilinear}
    \vspace{-\belowdisplayskip}
\end{figure}

To model the vehicle on roads with different shapes, we use a curvilinear coordinate system~\cite{lot2014curvilinear}, as illustrated in Fig.~\ref{fig:curvilinear} for a curved route shown by the solid black line. The advantage of using this coordinate system is that it is simpler to impose the path
constraints that force the vehicle to run within boundaries, i.e., withing reasonable distance from the road's centerline. The curvature of the path is represented by the curvature function $\tau(\gamma)$, where its magnitude is the reciprocal of the instantaneous radius at point $\gamma$ along the path, and the sign of $\tau(\gamma)$ depends on the direction of the curvature. In particular, for left turning curves, $\tau(\gamma)$ is negative and vice versa, $\tau(\gamma)$ is positive for right turning curves. The position of the vehicle along the path can be then traced by the below system of differential equations:
\begin{align}
    \frac{ds}{dt} &= \frac{v\cos(\beta) - u\sin(\beta)}{1+n\tau(s)}\label{eq:dsdt}\\
    \frac{dn}{dt} &= u\cos(\beta)+v\sin(\beta)\label{eq:dndt}\\
    \frac{d\beta}{dt} &= \delta - \tau(s)\frac{ds}{dt}\label{eq:dbetadt},
\end{align}
where $s$ is the distance of the vehicle along the path, $n$ is the lateral distance of the vehicle to the path, and $\beta$ is the difference between the angle of the longitudinal velocity vector of the vehicle and the tangent vector of the path at $s$ (see Fig.~\ref{fig:curvilinear} for visualization of the parameters).

To model the vehicle tire emission, we take after the model used previously in~\cite{singer2022influence}, in which they assume a strong correlation between the longitudinal force applied on the tires and their emission.
In particular, the number of tire particles generated from a tire is as follow:
\begin{equation}
    PN(F_{v}, p) = p_2F_v^2 + p_1 F_v + p_0,\label{eq:emission}
\end{equation}
where $F_v$ is the longitudinal force of the tire, and $p = \{p_2,p_1,p_0\}$ are fitting parameters. In this work, we obtain $p$ from fitting the measured data from~\cite{foitzik2018investigation} to a quadratic model, described fully  in~\cref{sec:res}.

\section{Methodology}
\label{sec:method}
Our proposal for lowering tire wear particle emission is based on the idea of dual-profile axes, i.e., using different tire profiles on the front (low-traction low-emission \enquote{hard} tires) and rear (high-traction high-emission \enquote{soft} tires) axes of the vehicle. The approach is to complement this hardware implementation with a control technology that corrects for the performance and dynamic differences resulting from the uneven tire traction. Our methodology, presented in this section, includes theoretical derivations based on the vehicle dynamics model in~\cref{sec:pre} and the application of a control algorithm. The proposed system aims to optimally divide the torque between vehicle axes and make small corrections to the driver's steering input to minimize the tire emissions and simultaneously match the driving experience to that of a vehicle running on all high-performing soft tires.

In the remainder of this work, we refer to the vehicle with all identical tires as the \emph{base} vehicle and the vehicle with different tire profiles as the \emph{low-wear} vehicle.
\subsection{The impact of hard tires on performance}
\label{sec:limit}
In this section, we discuss the performance difference that arises when swapping both wheels on an axis to a harder tire profile. Since the harder tire has lower peak forces compared to the base tire, it is no surprise that the low-wear vehicle cannot exactly replicate the performance of the base vehicle. However, this decrease in performance can become insignificant and only noticeable under extreme conditions under appropriate control, which we highlight in this section.

First, we addressed the lower peak longitudinal force in the low-wear vehicle. This results in slower acceleration and deceleration in extreme cases.
Ignoring the comfort limits of the driver, i.e., there are no limits on the acceleration and jerk of the vehicle, we can work out the approximate stopping distance $d$ via the following formulation from~\cite{HPWizard}:
\begin{equation}
    \label{eq:d}
    e^{Kgd}-1 =\frac{Kv^2}{2(\mu_{\max}+f_r)} \Leftrightarrow d = \frac{\ln\brac{Kv^2+2\brac{\mu_{\max}+f_r}}-\ln\brac{2\brac{\mu_{\max}+f_r}}}{Kg} \approx \frac{v^2}{2(\mu_{\max}+f_r) g},
\end{equation}
where $f_r = 0.013$ is the rolling resistance coefficient, and $g=9.81$~ms$^{-2}$ is the gravitational constant, $\mu_{\max}$ is the peak longitudinal friction coefficient of the vehicle. The constant $K$, for any road vehicle, is defined as: 
\begin{align}
    K = \frac{\rho C_DA}{mg},\label{eq:K}
\end{align}
where $\rho \approx 1.225~\text{kgm}^{-3}$ is the atmospheric air density, and $C_D\cdot A$ is the vehicle's drag area. The drag area is the frontal surface area of the vehicle that resists their forward motion, which is around $0.74 ~\text{m}^2$ for an average vehicle. So, for a typical vehicle with a mass of $m=1500$ kg, $K$ can be approximated as $6.16 \cdot 10^{-4} ~\text{s}^2\text{m}^{-2}$.
The peak friction coefficient of the vehicle is:
\begin{equation}
    \displaystyle\mu_{\max} = \frac{1}{l_r+l_f}\brac{l_r\max_{\lambda_f}\cM_{v}^{(f)}(\lambda_f) + l_r\max_{\lambda_r}\cM_{v}^{(r)}(\lambda_r)}
\end{equation}
where $\cM_{v}^{(f)}$ is the front tire model and $\cM_{v}^{(r)}$ is the rear tire model.
Assume that the distance from the center of gravity of the vehicle to its front axis and the distance from the center of gravity of the vehicle to its rear axis are approximately equal. When substituting the front axis with a harder tire with $x\%$ lower peak forces, the total peak friction coefficient of the vehicle is reduced by $(x/2)\%$ and the stopping distance increases by $(x/2)\%$. Let us consider the case where the soft tire has a peak friction coefficient between $0.5$ (wet asphalt) and $0.9$ (dry asphalt)~\cite{dimitrakopoulos2022cognitive}, and the hard tire has $20\%$ less peak friction coefficient compared to the soft tire. Our setup with a soft tire axis and a hard tire axis would have the peak friction coefficient between $0.45$ (wet asphalt) and $0.81$ (dry asphalt)~\cite{dimitrakopoulos2022cognitive}. This results in a $10\%$ increase in the stopping distance. If the vehicle moves at $30,60,120~\text{kmh}^\inv$, the stopping distance in dry condition of the low-wear vehicle will increase by approximately $0.4,1.7,7$ m, while in wet condition, the stopping distance will increase approximately by $0.8,3.14,12.6$ m.

As for the lower lateral force, when driving on a curve, it is subject to a centrifugal force of $mv^2r^\inv$, where $r$ is the turning radius.  
As long as the peak lateral force of the vehicle is greater than this centrifugal force, the vehicle is able to maintain the trajectory. 
Since we want the yaw acceleration to be approximately zero when the vehicle is turning at a steady state, the peak lateral force is approximately divided equally between the axes, which means the peak lateral acceleration of the vehicle can be calculated as:
\begin{equation}
    \displaystyle F_{u,\max} = mg\min\brac{\max_{\alpha_f}\cM_{v}^{(f)}(\alpha_f),\max_{\alpha_r}\cM_{v}^{(r)}(\alpha_r)}
\end{equation}
For a typical tire operating on concrete and asphalt roads in both wet and dry conditions, where $0.9\geq \max_{\alpha}\cM_{u}(\alpha)\geq 0.5$~\cite{dimitrakopoulos2022cognitive}, this peak lateral acceleration ranges from approximately $4.9~\text{ms}^{-2}$ to $8.8~\text{ms}^{-2}$. With the front tire substituted with a harder tire with $20\%$ lower peak lateral force, the peak lateral acceleration is reduced by $20\%$, resulting in a range of approximately $3.9~\text{ms}^{-2}$ to $7~\text{ms}^{-2}$.

However, note that for a typical driver, the comfortable maximum lateral acceleration is approximately $5.6~\text{ms}^{-2}$~\cite{bae2019toward}, so the lowered range of lateral acceleration does not impact the vehicle's performance in comfortable driving scenarios. Furthermore, per National Highways of the United Kingdom~\cite{dmrb}, the desirable minimum radius for a curved road is set such that the maximum lateral acceleration is much lower at approximately $1.09~\text{ms}^{-2}$. 
Therefore, unless the road is designed with a turning radius several times smaller than the recommended value or the driving conditions are so severe that friction is drastically reduced, replacing the front tire with a harder one does not significantly impair lateral maneuverability under normal conditions.

\subsection{Dynamics analysis}
\label{sec:scheme}
In the previous section, we have discussed the extreme scenarios where the low-wear vehicle can not fully replicate the dynamics of the base vehicle. In this section, we focus on the more likely cases where the low-wear vehicle is not operating at its limits and hence, as shown here, can be controlled to behave the same  the base vehicle given equal driver's inputs.
Here, we mathematically formulate this problem and propose a simple control algorithm that compensates for the differences in tire characteristics while also minimizing the tire emissions. The proposed method follows these steps:
\begin{itemize}
    \item First, we mathematically demonstrate that harder tires in the low-wear vehicle must be on the front axis, so that maintaining the same dynamics as the base vehicle would be possible. (Section \ref{sec:selectaxis})
    \item Then, we show that, given the soft and hard tire models, along with the driver’s steering input and sensor measurements of the low-wear vehicle’s dynamics, the base vehicle's dynamics under the same driver inputs can be predicted in terms of tire forces. (Section \ref{sec:baseforces}) 
    \item Finally, we formulate conditions for the longitudinal tire forces and the steering angle for the low-wear vehicle that equalize the dynamic response to that of the base vehicle given the same driver's input. Then, we proposed a simple search algorithm to find the optimal forces and steering angles that minimize the tire emission. (Section \ref{sec:optsteer})
\end{itemize}

Let us by adding a hat ( $\hat{}$ ) over the notations associated with the base vehicle differentiate them from the notations for the low-wear vehicle; e.g., $\ol{v}$ denotes the longitudinal velocity of the base vehicle, while $v$ denotes the longitudinal velocity of the low-wear vehicle.
First, we need to assume that the driver inputs to the low-wear vehicle are given to the base vehicle without any changes, i.e.:
\begin{equation}
    \sigma = \ol{\sigma},\,\ol {v} = v,\,\ol{u} = u,\,
    \ol{\delta} = \delta,\,\label{eq:uvd-equal}
\end{equation}
where $\sigma,u,v,\delta$ are the steering angle, longitudinal velocity, lateral velocity, and yaw rate, respectively. The driver directly decides the steering angle and longitudinal velocity, and the lateral velocity and yaw rate are dependent on the driver's inputs. For the two vehicles to have the same dynamics, the sum of forces along the vehicles longitudinal and lateral axis need to be the same, this means the following needs to be true at all times:
\begin{gather}
    \frac{d\ol{v}}{dt} = \frac{dv}{dt},\,\frac{d\ol{u}}{dt} = \frac{du}{dt},\,\frac{d\ol{\delta}}{dt} = \frac{d\delta}{dt}\label{eq:duvd-equal}.
\end{gather}

As it will be explained shortly, the proposed control system adjusts the steering angle by calculating a correction $\Delta\sigma$ so that the final steering $\sigma+\Delta\sigma$ would be applied to the low-wear vehicle (aiming at both replicating the base dynamics and optimizing emissions).

\subsubsection{Selecting the axis for tire substitution}\label{sec:selectaxis}
Given the same driver inputs, for the lateral acceleration to be equal between the base and the low-wear vehicles, from~\eqref{eq:duvd-equal} and by using~\cref{eq:dudt}, the difference between the rear tire forces of the two vehicles can be express as:
\begin{align}
    &\frac{d\ol{u}}{dt} = \frac{du}{dt}\nonumber\\
    \Leftrightarrow&\frac{1}{m}\brac{\ol{F}_{u,r} + \ol{F}_{v,f}\sin(\ol{\sigma}) + \ol{F}_{u,f}\cos(\ol{\sigma})}-\ol{v} \ol{\delta} = \frac{1}{m}\brac{F_{u,r} + F_{v,f}\sin(\sigma +\Delta\sigma) + F_{u,f}\cos(\sigma +\Delta\sigma)}-v \delta\nonumber\\
    \Leftrightarrow &\ol{F}_{u,r} - F_{u,r} =  \brac{F_{v,f}\sin(\sigma + \Delta\sigma) + F_{u,f}\cos(\sigma + \Delta\sigma)} - \brac{\ol{F}_{v,f}\sin(\ol{\sigma}) + \ol{F}_{u,f}\cos(\ol{\sigma})}
    \label{eq:du-equal}.
\end{align}
Now, similarly as above, for the yaw acceleration $d\delta/dt$ of both vehicle to be equal, using~\cref{eq:dpsidt2,eq:duvd-equal}, the difference between the rear tire forces of the two vehicles can be express as:
\begin{align}
    &\frac{d\ol{\delta}}{dt} = \frac{d\delta}{dt}\nonumber\\
    \Leftrightarrow &\frac{1}{I_z}\brac{-\ol{F}_{u,r}l_r + l_f\brac{\ol{F}_{v,f}\sin(\ol{\sigma}) + \ol{F}_{u,f}\cos(\ol{\sigma}) }} = \frac{1}{I_z}\brac{-F_{u,r}l_r + l_f\brac{F_{v,f}\sin(\sigma +\Delta\sigma) + F_{u,f}\cos(\sigma+\Delta\sigma) }}\nonumber \\
    \Leftrightarrow & \ol{F}_{u,r} - F_{u,r} = -\frac{l_f}{l_r}\brac{\brac{F_{v,f}\sin(\sigma+\Delta\sigma) + F_{u,f}\cos(\sigma+\Delta\sigma)} - \brac{\ol{F}_{v,f}\sin(\ol{\sigma}) + \ol{F}_{u,f}\cos(\ol{\sigma}) }}
    \label{eq:ddelta-equal},
\end{align}
Substitute~\cref{eq:du-equal} into~\cref{eq:ddelta-equal}, we have
\begin{align}
    \ol{F}_{u,r} - F_{u,r} = -\frac{l_f}{l_r}\brac{\ol{F}_{u,r} - F_{u,r}}
\end{align}
Since $l_f/l_r$ is nonzero and positive, it indicates that $\ol{F}_{u,r} - F_{u,r} = 0$, i.e., the difference between the rear tire forces of the two vehicle must be zero. Furthermore, the lateral rear tire force is a function of $u$, $v$, and $\delta$, which are kept constant across the vehicles. Hence, for $\ol{F}_{u,r}$ and $F_{u,r}$ to be equal, they must have the same tire model. This is not the case for the front axis, due to the fact that the steering angle is key in determining the tire forces and thus can be leveraged to equalize the forces on different tire profiles.
In conclusion, to equalize the dynamics between the base and low-wear vehicle, they need to have the same tire profile in the rear axis, thus, the rear tire should remain soft in the low-wear vehicle, while the front tire can be replaced with a harder tire profile.

\subsubsection{Modeling the base vehicle dynamics}\label{sec:baseforces}
With the front axis selected for the low-friction (hard) tires, we show that, given the tire characteristics and the necessary sensor data, the low-wear vehicle can be controlled to perform approximately the same as the base vehicle under the same driver inputs. 
The proposed control system in the low-wear vehicle does this in two steps: i) deriving the appropriate forces and a steering correction that mimics the predicted dynamics of the base vehicle (given the same inputs) in the low-wear vehicle with minimal emissions ii) the calculated force is then given as a reference to a torque controller, which controls the engine torque to generate the requested forces. Here, we formulate the first step of the control mechanism. 

To approximate the tire forces of the base vehicle, we take advantage of the equal dynamical responses of the two vehicles to the same driver input, i.e., lateral velocity, longitudinal velocity, yaw rate, and their time derivatives are the same at all times. 
From~\cref{eq:alphaf,eq:alphar,eq:Fu}, the lateral force of the front and rear wheel of the base vehicle are
\begin{align}
    \ol{F}_{u,f} &= \frac{mgl_r}{l_r+l_f}\cM_{u}(\ol{\alpha}_f)= \frac{mgl_r}{l_r+l_f}\cM_{u}^{(s)}\brac{\ol{\sigma}-\frac{\ol{u}}{\ol{v}}-\frac{l_f\ol{\delta}}{\ol{v}}} \label{eq:Fhatuf}\\
    \ol{F}_{u,r} &= \frac{mgl_f}{l_r+l_f}\cM_{\ol{u}}(\ol{\alpha}_r)= \frac{mgl_f}{l_r+l_f}\cM_{u}^{(s)}\brac{-\frac{\ol{u}}{\ol{v}}+\frac{l_r\ol{\delta}}{\ol{v}}}\label{eq:Fhatur}.
\end{align}
We assumed the the soft-tire model $\cM_{u}^{(s)}(\cdot)$ is available. The parameters $\ol{u}$ (longitudinal velocity), $\ol{v}$ (lateral velocity), $\ol{\delta}$ (yaw rate) can be replaced in the above with $u$, $v$, and $\delta$ monitored in the low-wear vehicle, assuming the same movement in both vehicles.
By measuring the driver steering input in the low-wear vehicle $\sigma$, and assuming it is given to the base vehicle $\ol{\sigma}=\sigma$, the lateral forces of the base vehicle can be exactly predicted using \cref{eq:Fhatuf} and \cref{eq:Fhatur}.
Furthermore, as we monitor the time series of $u$ and $v$, we can also approximate their derivatives, $dv/dt$ and $du/dt$, which have the same value as that of the base vehicle, i.e., $d\ol{v}/dt = dv/dt$ and $d\ol{u}/dt = du/dt$.
The longitudinal forces for the base vehicle can then be calculated by solving the following system of equations derived from~\cref{eq:dvdt,eq:dudt}:
\begin{align}
&\begin{cases}
    \frac{d\ol{v}}{dt} = \frac{1}{m}\brac{\ol{F}_{v,r} + \ol{F}_{v,f}\cos(\ol{\sigma}) - \ol{F}_{u,f}\sin(\ol{\sigma}) - C_d\ol{v}^2}-\ol{u}\ol{\delta}\\
    \frac{d\ol{u}}{dt} = \frac{1}{m}\brac{\ol{F}_{u,r} + \ol{F}_{u,f}\cos(\ol{\sigma}) + \ol{F}_{v,f}\sin(\ol{\sigma})} - \ol{v}\ol{\delta}
\end{cases}\\
\Leftrightarrow&\begin{cases}
    \ol{F}_{v,r} + \ol{F}_{v,f}\cos(\ol{\sigma}) = m\frac{d\ol{v}}{dt} - m\ol{u}\ol{\delta} + C_d\ol{v}^2 + \ol{F}_{u,f}\sin(\ol{\sigma}) \\ 
 \ol{F}_{v,f}\sin(\ol{\sigma}) = m\frac{d\ol{u}}{dt} + m\ol{v}\ol{\delta} - \ol{F}_{u,r} - \ol{F}_{u,f}\cos(\ol{\sigma}).
\end{cases}\\
\Leftrightarrow& \begin{bmatrix}
    1& \cos(\ol{\sigma})\\
    0 & \sin(\ol{\sigma})
\end{bmatrix}\begin{bmatrix}
    \ol{F}_{v,r}\\
    \ol{F}_{v,f}
\end{bmatrix} = \begin{bmatrix}
    m\frac{d\ol{v}}{dt} - m\ol{u}\ol{\delta} + C_d\ol{v}^2 + \ol{F}_{u,f}\sin(\ol{\sigma})\\
    m\frac{d\ol{u}}{dt} + m\ol{v}\ol{\delta} - \ol{F}_{u,r} - \ol{F}_{u,f}\cos(\ol{\sigma})
\end{bmatrix}\label{eq:FvrFvf}.
\end{align}
As all quantities on the right of~\cref{eq:FvrFvf} can be approximated by this point, we can continue to approximate the longitudinal tire forces of the base vehicle $\ol{F}_{v,r}$ and $\ol{F}_{v,f}$ as long as the leftmost matrix has non-zero determinant, i.e., $\sin(\ol{\sigma}) \neq 0 \Leftrightarrow \ol{\sigma} \neq 0$. This is the case where there is no steering input from the user, in which case, as long as the sum of the longitudinal forces is the same, the vehicles would have the same dynamics. In the following section, these two cases are handled separately. In the case where there is no steering, we find the appropriate tire forces, and in the case where there is steering, we find both the appropriate tire forces and the adjusted steering angle.

\subsubsection{Optimizing steering angle and tire forces}
\label{sec:optsteer}
In this section, we determine the appropriate tire forces and steering adjustments for the low-wear vehicle to replicate the dynamics of the base vehicle under the same driving inputs while also minimizing tire emissions. By approximating the base vehicle's tire forces and equalizing the sum of forces along both the longitudinal and lateral axes of the two vehicles, we formulate a set of conditions for the forces and steering angle of the low-wear vehicle to ensure the equivalence of driving performance between the two vehicles.
From these conditions, we can select a set of tire forces and steering adjustments that minimize the tire emission. This is achieved by reformulating the emission formulation in~\cref{eq:emission} such that it depends solely on a single variable. This allows for efficiently finding an optimal solution either analytically or through a greedy searching algorithm. 

When the vehicle is under minimal steering ($\sigma\approx 0$), the steering angle, lateral forces, and lateral velocity are close to zero. From~\cref{eq:dvdt,eq:dudt,eq:dpsidt2} we have: 
\begin{gather}
    \dx{v}{t} = \frac{1}{m}\brac{F_{v,r} + F_{v,f}-C_dv^2},\nonumber \\
    \frac{du}{dt} = 0,\quad\frac{d\sigma}{dt} = 0,\quad\frac{d\delta}{dt} = 0.
\end{gather}%
It is seen that in the minimal steering scenario, as long as $F_{v,f}+F_{v,r} = mdv/dt+C_dv^2$, the longitudinal forces can be divided freely, and the low-wear vehicle's movement will be the same as the base vehicle. To find the optimal forces for tire emission, we can substitute $F_{v,r} = mdv/dt+C_dv^2-F_{v,f}$ into~\cref{eq:emission} and get
\begin{align} 
PN_{\text{total}} = PN\brac{F_{v,f},p^{(h)}} + PN\brac{mdv/dt+C_dv^2-F_{v,f},p^{(r)}},\label{eq:ttl-emission-2}
\end{align}where $p^{(s)}$ and $p^{(h)}$ are the emission coefficients for the soft and the hard tire, respectively. This reformulation leaves $F_{v,f}$ as the only free variables, for which the optimal value (minimizing the tire emissions) can be simply found and be used to find the optimal $F_{v,r}$.

As for the case when the steering input is nonzero, where~\cref{eq:FvrFvf} has a unique solution, we are able to individually predict the longitudinal tire forces of the base vehicle. 
This enables us to also individually calculate the respective tire forces for the low-wear vehicle such that the two vehicles have identical dynamics.
The forces to be put into the low-wear vehicle also depend on the new adjusted steering angle, as a bigger steering angle converts some of the vehicle's longitudinal acceleration into lateral acceleration, thus requiring more forces in the longitudinal direction compared to the base vehicle to compensate.
Due to this dependence on the adjusted steering angle, TWP emissions can be reformulated as a function that solely depends on the steering angle adjustment $\Delta\sigma$.
Using~\cref{eq:alphaf,eq:Fu}, the lateral force of the front wheel in the low-wear vehicle can be written as:
\begin{align}
    F_{u,f} &= \frac{mgl_r}{l_r+l_f}\cM_{u}^{(h)}(\alpha_f)= \frac{mgl_r}{l_r+l_f}\cM_{u}^{(h)}\brac{\sigma+\Delta\sigma-\frac{u}{v}-\frac{l_f\delta}{v}} \label{eq:Fuf}.
\end{align}
It was established earlier that for the vehicle to have the same driving performance,~\cref{eq:du-equal} must be true. Now, added on the fact that  the lateral rear tire forces are the same, i.e., $F_{u,r} = \ol{F}_{u,r}$, 
for the two vehicles to drive the same, $\Delta\sigma$ and $F_{v,f}$ must satisfy the following
\begin{align}&F_{v,f}\sin(\sigma+\Delta\sigma) + F_{u,f}\cos(\sigma+\Delta\sigma) = \ol{F}_{v,f}\sin(\sigma) + \ol{F}_{u,f}\cos(\sigma)\nonumber \\
\Leftrightarrow &F_{v,f} = \frac{\ol{F}_{v,f}\sin(\sigma) + \ol{F}_{u,f}\cos(\sigma) - F_{u,f}\cos(\sigma+\Delta\sigma)}{\sin(\sigma+\Delta\sigma)}\label{eq:opt-Fvf}
\end{align}
Next, for the two vehicles to have the same dynamics, $\Delta\sigma$ and $F_{v,r}$ must satisfy the following
\begin{align}
F_{v,r}  &= m\frac{dv}{dt} - mu\delta + C_dv^2 + F_{u,f}\sin(\sigma+\Delta\sigma) -F_{v,f}\cos(\sigma+\Delta\sigma)\label{eq:opt-Fuf}
\end{align}
From~\cref{eq:opt-Fuf,eq:opt-Fvf}, for both vehicles to have the same movement, both longitudinal forces $F_{v,f}$ and $F_{v,r}$ can be formulated as functions that solely depend on the steering adjustment angle $\Delta\sigma$. Thus, \cref{eq:opt-Fuf,eq:opt-Fvf} can be used to express the emission formula such that depends solely on $\Delta\sigma$.
\begin{equation}
    PN_{\text{total}} = PN\brac{F_{v,f},p^{(h)}} + PN\brac{F_{v,r},p^{(r)}}\label{eq:ttl-emission}
\end{equation}
where $p^{(s)}$ and $p^{(h)}$ are the emission coefficients for the soft and the hard tire, respectively.
Now, as $\Delta\sigma$ is the only free variable in this case, we can use a simple greedy local search algorithm to find a value of $\Delta\sigma$ that finds a local minimum for tire emissions. 
The local search algorithm aims to minimize tire emissions by iteratively adjusting the parameter $\Delta\sigma$. At each step, the estimated particle number is estimated at $\Delta\sigma - \varepsilon$ and $\Delta\sigma + \varepsilon$. If either of these values results in a lower particle number than at $\Delta\sigma$, the algorithm updates $\Delta\sigma$ to that value. If both neighboring estimates yield higher particle numbers, then $\Delta\sigma$ is close to a local minimum. In this case, the search range $\varepsilon$ is reduced and the process continues until $\varepsilon$ becomes sufficiently small, ensuring convergence to a local minimum. The algorithm is described in detail in~\cref{alg:search}.
Now, given the optimal adjusted steering angle, we can find the associated optimal longitudinal tire forces using~\cref{eq:opt-Fuf,eq:opt-Fvf}.

\begin{algorithm}[t]
\SetKwInOut{Input}{input}  
\Input{Measurement data: $\displaystyle\sigma,u,v,du/dt,dv/dt,\delta$; 
Tire models: $\cM^{(s)}(\cdot), \cM^{(h)}(\cdot)$; 
Tire emission paramter: $p^{(s)}, p^{(h)}$;
Search precision: $\varepsilon_{\min}$;
}
\SetKwInOut{Output}{output}  
\Output{ Steering adjustment angle: $\Delta\sigma$;}
\hrulefill

Initialize $\Delta\sigma \gets 0$; $i \gets 1$; $\varepsilon \gets 10^{-2}$\;
\While{$\varepsilon \geq \varepsilon_{\min}$}{
    $\displaystyle \sigma_{\text{left}} \gets \sigma + \Delta\sigma-\varepsilon$\;
    $\displaystyle \sigma_{\text{right}} \gets\sigma + \Delta\sigma+\varepsilon$\;
    $\displaystyle f \gets \text{estimatePN}(\sigma + \Delta\sigma,\sigma,u,v,du/dt,dv/dt,\delta)$\;
    $\displaystyle f_{\text{right}} \gets \text{estimatePN}(\sigma_{\text{right}},\sigma,u,v,du/dt,dv/dt,\delta)$\;
    $\displaystyle f_{\text{left}} \gets \text{estimatePN}(\sigma_{\text{left}},\sigma,u,v,du/dt,dv/dt,\delta)$\;
    \uIf{$f_{\text{right}} < f$}{%
        $\Delta\sigma \gets \sigma_{\text{right}}-\sigma $\;
    }\uElseIf{$f_{\text{left}} < f$}{%
        $\Delta\sigma \gets\sigma_{\text{left}}-\sigma $\;
    }\Else{
        $\varepsilon \gets \varepsilon/2$\;
    }
}
\Return{$\sigma$}\;
\SetKwProg{Fn}{function}{}{end}
\Fn{\textup{estimatePN}($\displaystyle \Delta\sigma,\sigma,u,v,du/dt,dv/dt,\delta$)}{
    \vspace{1mm}
    $\displaystyle\ol{F}_{u,f} \gets mgl_r\cM^{(s)}(\sigma-u/v-l_f\delta/v)/(l_r+l_f)$\tcc*[r]{~\cref{eq:Fhatuf}}
    $\displaystyle\ol{F}_{u,r} \gets mgl_r\cM^{(s)}(-u/v+l_r\delta/v)/(l_r+l_f)$\tcc*[r]{~\cref{eq:Fhatur}}
    $\displaystyle\ol{F}_{v,f} \gets m(du/dt)+mv\delta - \ol{F}_{u,r} - \ol{F}_{u,f}\cos(\sigma)$\tcc*[r]{~\cref{eq:FvrFvf}}
    $\displaystyle F_{v,f} \gets \ol{F}_{v,f}\sin(\sigma) + \ol{F}_{u,f}\cos(\sigma) - F_{u,f}\cos(\sigma +\Delta\sigma)/\sin(\sigma + \Delta\sigma)$\tcc*[r]{~\cref{eq:opt-Fuf}}
    $\displaystyle F_{v,r} \gets m(dv/dt)-mu\delta + C_dv^2 + F_{u,f}\sin(\sigma + \Delta\sigma) - F_{v,f}\cos(\sigma +\Delta\sigma)$\tcc*[r]{~\cref{eq:opt-Fvf}}
    \Return{$\displaystyle PN(F_{v,f},p^{(h)}) + PN(F_{v,r},p^{(s)})$}
}
\caption{Greedy local search algorithm to estimate the adjusted steering angle ($\sigma + \Delta\sigma$)}
\label{alg:search}
\end{algorithm}

As we do not directly control the tire force, the force calculated here is used as a reference. The vehicle controller finds the appropriate torque to minimize the error between the actual forces and the reference forces. For our simulation, after calculating the reference force signals using the driver input, we simulate the output of the torque controller by solving an optimal control problem. The formulation of the optimal control problem is presented in~\cref{sec:ssm}.

\begin{figure}[t]
    \centering
    \includegraphics[width=\linewidth]{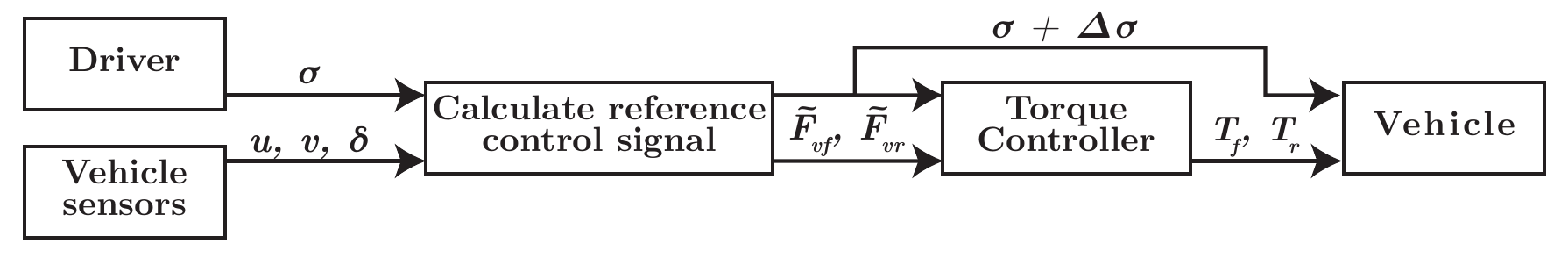}
    \caption{\textbf{Flow chart of the control algorithm.} The driver's steering input and the steering adjustment angle are denoted by $\sigma$ and $\Delta\sigma$, respectively. The vehicle movement is described by longitudinal velocity ($u$), lateral velocity ($v$), and yaw rate ($\delta$). The reference force signals for the front and rear tires are denoted by $\Tilde{F}_{v,f}$ and $\Tilde{F}_{v,r}$, respectively. The input torques to the front and rear wheels are denoted by $T_f$ and $T_r$, respectively.}
    \label{fig:flowchart}
\end{figure}

We end this section with a summary of the entire control process.
First, the steering angle input of the driver $\sigma$ and vehicle longitudinal velocity $v$, lateral velocity $u$, and yaw rate $\delta$ is collected. The data is passed to \cref{alg:search} which calculates the steering adjustment angle $\Delta\sigma$, and reference force signals for the front and rear tire $\Tilde{F}_{v,f}$ and $\Tilde{F}_{v,r}$. A torque controller then determines the appropriate torque input to the wheels to minimize the error between the references and actual forces. Finally, the adjusted steering angle ($\sigma +\Delta\sigma$) and the calculated torque ($T_f$ and $T_r$) are applied to the vehicle. The entire control process is illustrated in~\cref{fig:flowchart}.

\subsection{Optimal control problem}
\label{sec:ssm}

The simulation in this work is based on a modified state-space representation, where the vehicle dynamics are represented at intervals along a predetermined path rather than at time intervals, as in conventional state-space representations of dynamic systems. Our dynamics system representation is similar to the approach used by Papaioannou et al.~\cite{papaioannou2022optimal,papaioannou2024reducing} and Lot et al.~\cite{lot2014curvilinear}.
In this section, we formulate the state-space representation of the vehicle along a predetermined path, and later in~\cref{sec:simres}, the state-space model is used to compare the dynamics between the base vehicle and the low-wear vehicle with the adjusted force and steering. 

First, we discretize a path $\cP$ of length $|\cP|$ into $d$ equal sections of length $\Delta s = |\cP|d^\inv$. We use $s_i$ to denote the point on the path that makes an arc length of $i\Delta s$ with the starting point of $\cP$. 
Let us define a state vector 
\[\cQ(i) =[t(i),n(i),\beta(i),u(i),v(i),\delta(i),\omega_f(i),\omega_r(i)]^T,\]
which describes the state of the vehicle at step $i$, we use $\cQ$ to denote the ordered set of all states $\cQ = [\cQ(0),\dots,\cQ(d)]$.
Each entry in vector $\cQ(i)$ is the value of a function at step $i$ (defined in~\cref{sec:pre}), with the exception of $t$, which is the function of elapsed time at step $i$. The ordinate $n(i)$ indicates the lateral distance of the vehicle from the point $s_i$ on the path, and $\beta(i)$ is the angle between the vehicle's heading to the tangent of the path at the point $s_i$. Lateral velocity $u(i)$, longitudinal velocity $v(i)$, and yaw rate $\delta(i)$, determine the vehicle dynamics with respect to its reference frame at step $i$. The angular velocity of the two-wheel axes at step $i$ is indicated by $\omega_f(i)$ and $\omega_r(i)$. 

We also define a state derivative vector 
\[\cD(i) =\sbrac{\frac{dt}{ds}(i),\dx{n}{s}(i),\dx{\beta}{s}(i),\dx{u}{s}(i),\dx{v}{s}(i),\dx{\delta}{s}(i),\dx{\omega_f}{s}(i),\dx{\omega_r}{s}(i)}^T,\]
where each entry of $\cD(i)$ is a derivative of a function w.r.t. the path at step $i$ (or when the closest point on the path to the vehicle is $s_i$). To calculate the derivatives over the path length, we use the chain rule with time derivatives of the relevant function and $\dx{s}{t}(i)$ (from~\cref{eq:dsdt}). E.g., $\dx{n}{s}(i) = \dx{n}{t}(i)/\dx{s}{t}(i)$.
See~\Cref{tab:state-func} for a short description of all the functions and the definition of their time derivatives
Finally, we define an input vector $\cU(i) = [T_f(i),T_r(i)]$, which describe the input at step $i$, we also use $\cU$ to denote the ordered set of all state $\cU = [\cU(0),\dots,\cU(d)]$. Within $\cU(i)$, $T_f(i)$ and $T_r(i)$ indicate the input torques for the front and rear wheels at step $i$, respectively.

\begin{table}[t]
 \caption{Summary of the vehicle state functions and reference to the calculation of the derivatives.}
  \centering
  \begin{tabular}{llll}
    \toprule               
    Function & Description & Unit & Time derivative \\
    \midrule
    $t$ & Elapsed time &s & 1\\
    $n$ & Ordinate in the curvilinear coordinate &m & \cref{eq:dndt}\\
    $\beta$ & Heading angle in curvilinear coordinate &rad &\cref{eq:dbetadt}\\
    $u$ & Lateral velocity &ms$^{\inv}$ & \cref{eq:dudt}    \\
    $v$ & Longitudinal velocity &ms$^{\inv}$ & \cref{eq:dvdt}      \\
    $\delta$ &Yaw rate &rads$^{\inv}$      & \cref{eq:dpsidt2}  \\
    $\omega_f$ & Front wheel angular velocity &rads$^{\inv}$ & \cref{eq:domegadt}\\
    $\omega_r$ & Rear wheel angular velocity &rads$^{\inv}$ & \cref{eq:domegadt}\\
    \bottomrule
  \end{tabular}
  \label{tab:state-func}
\end{table}

To determine the next system state $\cQ(i+1)$ from the current state $\cQ(i)$, given that $\Delta s$ is sufficiently small, we can use the linear approximation:
\begin{equation}
    \cQ(i+1) = \cQ(i) + \Delta s \cdot \cD(i)~~~\text{for}~i = 0,\dots,d-1.    
\end{equation}

Using the state space representation, we can formulate the following optimal control problem:
\begin{align}
    \begin{array}{rl}
    \displaystyle \min_{\cQ,\cU}& \cJ(\cQ,\cU)\\
    \textup{s.t.} &\displaystyle \cQ(i) = \cQ(i-1) + \Delta s\cdot \cD(i-1)\\
    & g\brac{\cQ(i),\cU(i)} \leq 0 \;\text{for}\;i = 1,\dots,d,\label{eq:opti}
\end{array}
\end{align}
where $\cJ(\cdot)$ is an objective function, and $g(\cdot)$ are the constraint functions. As previously mentioned, given reference force signals, we find the torque input that minimizes the error between the reference and the actual force generated. Let $\Tilde{F}_{v,f}(i)$ and $\Tilde{F}_{v,r}(i)$ be the reference signal for the longitudinal force of the front and rear tires at step $i$, respectively. The objective function $\cJ$ is hence:
\begin{equation}
    \cJ(\cQ,\cU) = \sum_{i=1}^d \brac{F_{v,f}(i) - \Tilde{F}_{v,f}(i) }^2+ \brac{F_{v,r}(i)  - \Tilde{F}_{v,r}(i) }^2
\end{equation}
The constraints of the optimal control problem are that the steering angle is fixed to the adjusted steering ($\sigma + \Delta\sigma$), and the vehicle's longitudinal speed is set as the driver input.

\section{Results}
\label{sec:res}

In this section, we perform present data analysis related to practical application of our proposal and then numerically assess the effectiveness of the proposed system involving different tire profiles on different axes and a controller that optimizes the performance and tire-wear emissions of the vehicle. In the following, first, the relation between tire emissions and traction is modeled based on real data from common tires in the market. We then use this model in the formulation of our proposed control system to perform optimization on torque division and steering correction to achieve the main two aims of the paper: i) minimizing the tire wear emissions, and ii) maintaining the performance and driving experience of the low-wear vehicle (running on different tires) to that of the base vehicle (running on all high-performing soft tires). The proof of concept is conducted using detailed numerical simulations of different scenarios and measuring the extent to which the above two aims are achieved with the proposed system.

\subsection{Tire wear versus traction}

We derive the relationship between wear and traction of the common tires using real data collected from publicly available resources. An important reference for tire performance data is the Uniform Tire Quality Grading (UTQG) system~\cite{UTQG2024}, established by the U.S. National Highway Traffic Safety Administration (NHTSA). Every passenger car tire sold in the United States is required to display its UTQG rating. The UTQG tire rating consists of three categories: treadwear, traction, and temperature. The treadwear rating, denoted by $w$, is a comparative measure in which each tire is tested alongside a benchmark tire on a vehicle driven in a convoy on a government test course. Tire wear is assessed during and after the test, compared to the benchmark tire, and recorded. For example, a tire with a treadwear rating of $w=200$ is expected to last twice as long as a tire rated $w=100$ under the same test conditions.

\begin{figure}[b]
    \centering
    \captionsetup{width=0.7\textwidth}
    \includegraphics[width=0.5\textwidth]{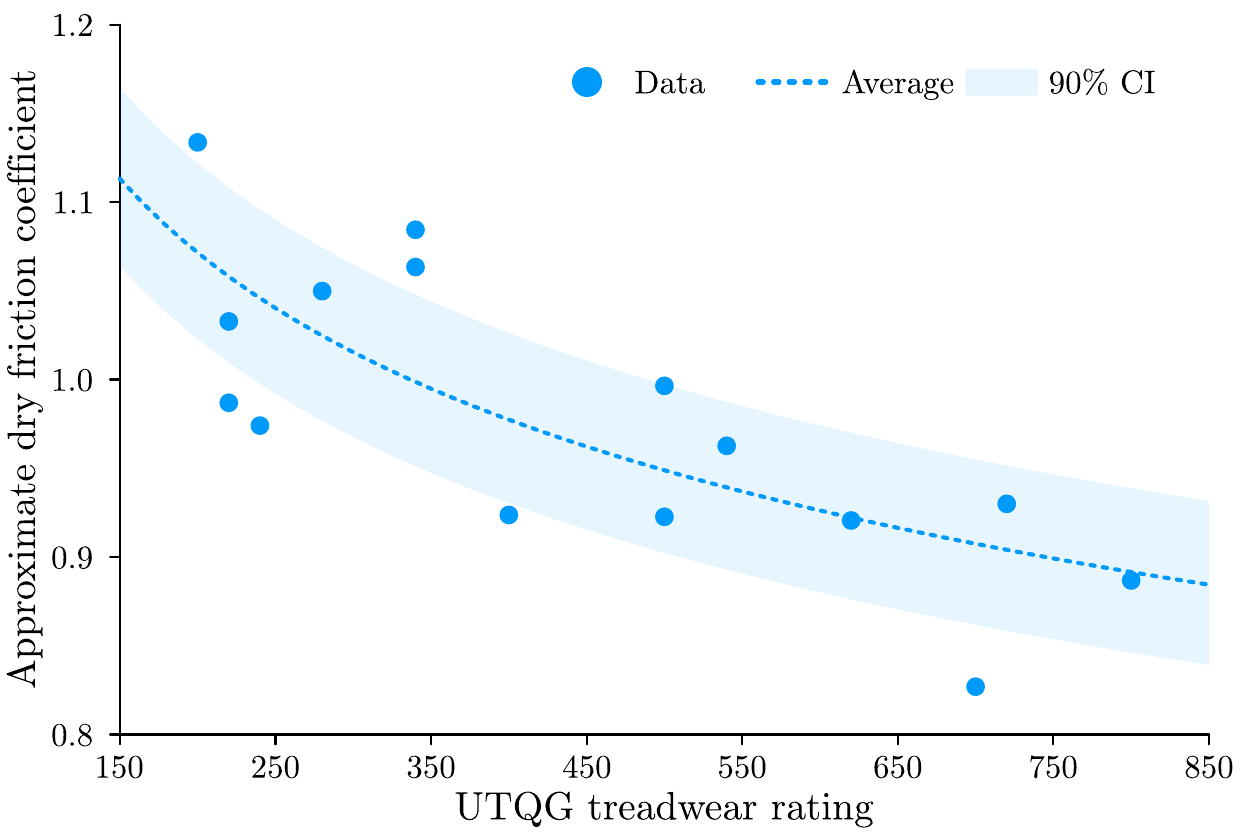}
    \caption{\textbf{Traction versus treadwear curve.} Approximated friction coefficient on dry asphalt against UTQG treadwear rating based on data from fifteen tires. Blue points are the collected data, the dashed line is the fit in Eq.~\eqref{eq:w-vs-mu}, and the shaded area marks the $90\%$ confidence interval (CI) for the fit.}
    \label{fig:tw-vs-tr}
    \vspace{-\belowdisplayskip}
\end{figure}%

Although traction is included in the UTQG rating, we do not use this data since the ratings are limited to four broad categories, providing only very rough approximations on actual traction values and failing to offer meaningful insight into the potential relationship between wear and traction of common tires. Instead, we use test data from Tire Rack~\cite{TRackTR}, which includes vehicle stopping distances from $v=50$ mph, allowing us to estimate the peak tire friction coefficient $\mu$ by rewriting Eq.~\eqref{eq:d}:
\begin{align}
    \mu = \frac{K \cdot v^2}{2(e^{Kgd}-1)}-f_r,\;K = 6.16 \cdot 10^{-4} ~\text{s}^2\text{m}^{-2},\label{eq:stop-dist}
\end{align}

Using data collected on fifteen tires from the 2024 UTQG report~\cite{UTQG2024}, along with corresponding stopping distance data from Tire Rack~\cite{TRackTR}, we model the relationship between the wear and traction of common tires. The relationship can be described with reasonable accuracy by the below equation:
\begin{equation}
\mu = 2.16 \cdot w^{-0.133}. \label{eq:w-vs-mu}
\end{equation}
The relationship between tire traction and treadwear has previously been quantified using the same model and reported to be $\mu = 2.25 \cdot w^{-0.15}$~\cite{HPWizard}, which is in agreement and very similar to the fitting  obtained here independently. Figure ~\ref{fig:tw-vs-tr} shows the collected data and the fitted model of Eq. \eqref{eq:w-vs-mu}. 

This result demonstrates that it is possible to select tires that significantly reduce emissions with only moderately compromised traction. For example, according to the fitted curve, a tire with a treadwear rating of $200$ has an average friction coefficient of $1.07$, while a tire with a treadwear rating of $800$ has a friction coefficient of $0.89$. This implies a $17\%$ reduction in traction in exchange for a $75\%$ decrease in emissions.
 
\subsection{Numerical Experiments}
The experiments involve simulations of two vehicles, the \emph{base} vehicle with all high-performing high-emission tires of the same profile, and our proposed \emph{low-wear} vehicle set up where the front axis tires are replaced with low-traction low-emission tires. Driving with both vehicles is then simulated in a number of scenarios with different trajectories and conditions. Our proposed methodology controls the torque division between axes and applies an adjusted steering angle in the low-wear vehicle. Our control scheme objective is to simultaneously minimize the tire wear emissions and minimize the difference in the driving performance of the low-wear vehicle to that of the base vehicle (so that the driving performance would be as if all four tires are high-performance soft tires). 

\subsubsection{Experimental setting}
\label{sec:setting}
\begin{figure}[b]
    \centering
    \captionsetup{width=1\textwidth}
    \includegraphics[width=0.9\linewidth]{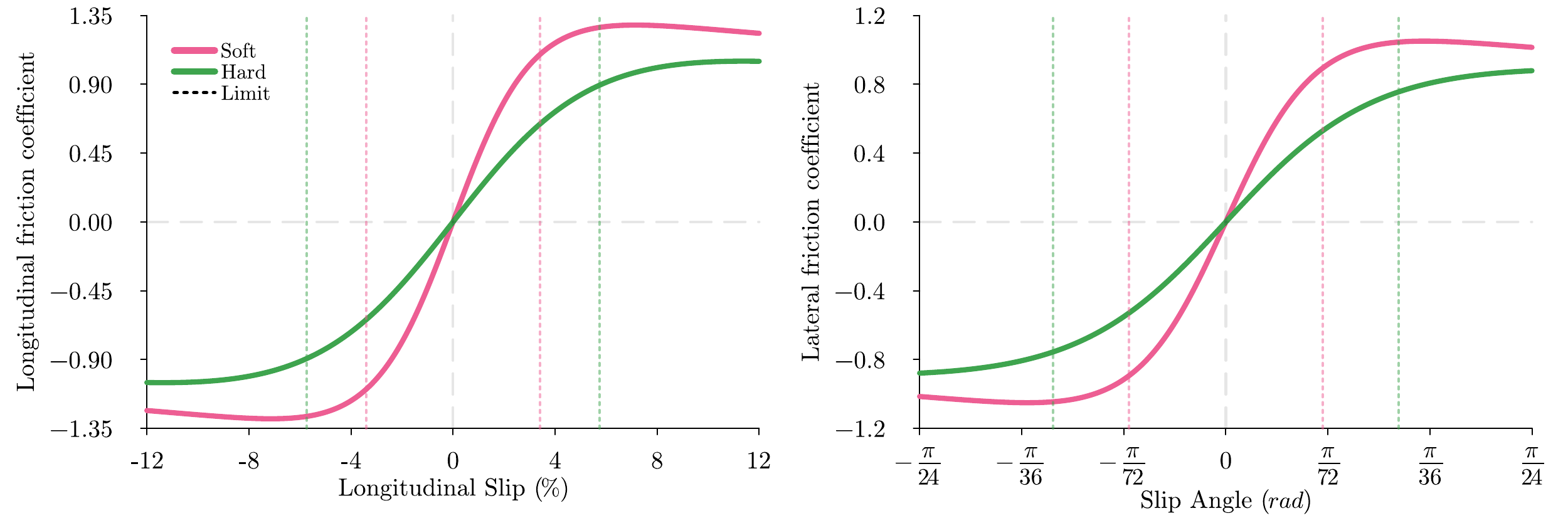}
    \caption{\textbf{Friction coefficient versus slip curves.} Longitudinal (left) and lateral (right) friction coefficient of the soft and hard tires used for simulation for the soft (red) tire and the hard (green) tire. The dotted line is the slip limits that the controller would impose on the tires}
    \label{fig:tirecurve}
    \vspace{-\belowdisplayskip}
\end{figure}

We simulate the dynamics for a vehicle with mass $m=1500$ kg, with a wheelbase of $2$ m, where each wheel axis is $1$ m from the center of mass. The air drag coefficient is fixed at $0.39$, and yaw axis inertia is set to $1800~\text{kgm}^2$. Each tire has a radius of $0.3$ m, and a wheel inertia of $0.8~\text{kgm}^2$.
For the tire model, we fit two built-in tire models from MATLAB's vehicle dynamic blockset plug-in, namely, \enquote{Light passenger car 205/60R15} and \enquote{Mid-size passenger car 235/45R18} to the Magic Formula \cite{pacejka2005tire}. Let us label the tire with the lower peak force as the \enquote{hard} tire and the one with the higher peak force as the \enquote{soft} tire. The so-called slip curves of these tires, describing the relationship between the friction coefficient and the slip, are inputs to our formulation of the vehicle dynamics. The slip curves of the soft and hard tires selected here are illustrated in Fig.~\ref{fig:tirecurve}, separately for longitudinal (left plot) and lateral (right plot) friction coefficient.

To simulate bad driving conditions (for example, driving on wet asphalt or having a pair of tires with worse performance), we also add a parameter $\zeta \in [0,1]$ and modify the tire model into $\zeta\cM(x)$, this proportionally reduces the peak tire force by $1-\zeta$. For normal driving conditions, $\zeta=1$, and for wet conditions, we set $\zeta = 0.5$, which reduces the peak tire force by $50\%$~\cite{dimitrakopoulos2022cognitive}.

To select the emission parameter for the tires, we utilize the quadratic emission model in~\cite{singer2022influence}. Then, for the different tires, we assume the hard (low-emission) tire emits a fixed fraction less than the soft (high-emission) tire in all operating points, i.e., $PN_h(F_v) = \rho\cdot PN_s(F_v)$ for any value of $F_v$. 

\begin{figure}[h]
    \centering
    \captionsetup{width=0.8\textwidth}
    \includegraphics[width=0.45\textwidth]{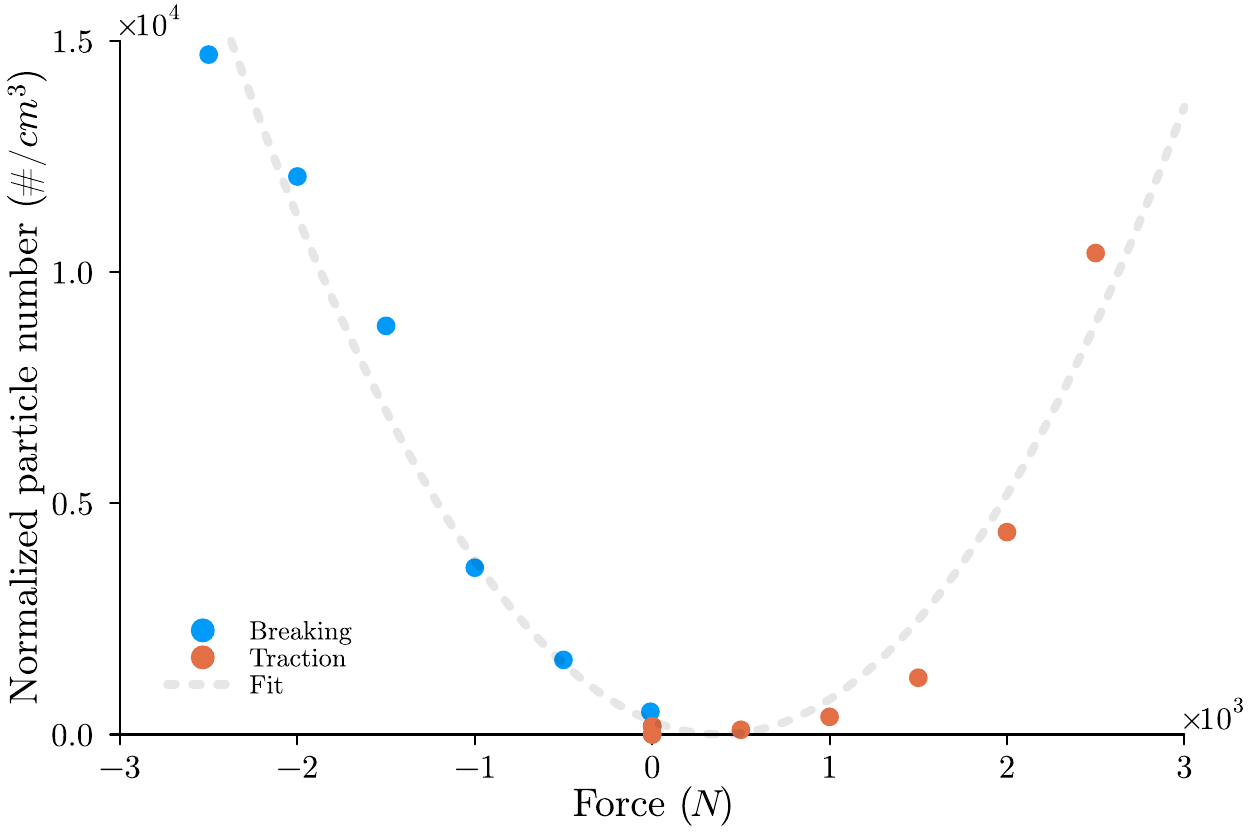}
    \caption{\textbf{Particle number versus longitudinal force.} The particle number data from~\cite{foitzik2018investigation} fitted to a single quadratic. Orange points denote the data from traction, and blue dots denote the data from braking. The fitted quadratic is displayed with the dotted line}
    \label{fig:emissionfit}
    \vspace{-\belowdisplayskip}
\end{figure}

To find the parameters $p$ for the emission model in~\cref{eq:emission}, we fit the emission data from both traction and braking data from Foitzig et al.~\cite{foitzik2018investigation} to a single quadratic, with the constraint that the emission is never negative. Solving a least square constrained optimization problem, gives us the fitting of $PN(F_v) =  1.98\cdot 10^{-3} \cdot F_v^2 - 1.50\cdot F_v+286.04$; the fit is very accurate with a coefficient of determination of $0.965$.
The data and the fitted quadratic are presented in~\cref{fig:emissionfit}, where the blue data points represent the particle count recorded during braking and the orange data points represent the particle count recorded during traction. 

The maximum traction of the hard tire taken from MATLAB is approximately $17\%$ less than that of the soft one. I.e., $\frac{\mu_h}{\mu_s} \approx 0.83$, where $\mu_h$ and $\mu_s$ are the approximate dry friction coefficient for the hard and soft tire, respectively. We can use the traction and treadwear relation in~\cref{eq:w-vs-mu} to calculate the emission scaling ratio $\rho$, as follows:
\begin{equation}
    \rho = \frac{w_h}{w_s} = \brac{\frac{\mu_h}{2.16}}^{7.52}\brac{\frac{\mu_s}{2.16}}^{-7.52} = \brac{\frac{\mu_h}{\mu_s}}^{7.52} = 0.83^{7.52} \approx \frac{1}{4}.
\end{equation}
where $w_h$ and $w_h$ are the hard and soft tire treadwear ratings, respectively.
Hence, for the soft tire, we use the parameter acquired from the fit previously, i.e., $p^{(s)} = \{1.98\cdot 10^{-3},-1.5,286.04\}$, and for the hard tire, we scale the fitted parameter by $\rho = 0.25$, and use $p^{(h)} = \{4.95\cdot 10^{-4},-10.375,71.51\}$.

For our simulations, we test three scenarios. First, an emergency braking scenario, where both vehicles attempt to come to a stop as quickly as possible. Then, for the second and third scenarios, we examine the vehicle in normal driving scenarios in straight line paths and in curved paths. In the two latter scenarios, we generate the trajectory of the base vehicle by solving the optimal control problem, where the objective is to complete the trajectory in the least amount of time while staying within predefined road boundaries. In particular, the vehicles are constrained to stay within a meter from the centerline of the defined path, i.e., 
\begin{equation}
    |n|\leq 1.
\end{equation} 
To generate a trajectory that is representative of that of a human driver, the vehicle is constrained to stay within the comfort limit of the driver. For this, we follow the \emph{Aggressive Driver} profile from the Occupant’s Preference Metric (OPM) in Bae et al.~\cite{bae2019toward,bae2020self}, which puts limits on the lateral and longitudinal acceleration and jerk of the vehicle as:
\begin{alignat}{5}
    -5.6 &\leq \frac{du}{dt}     &\leq 5.6 &,\quad & \frac{dv}{dt}     &\leq 3.07\\
    -2   &\leq \frac{d^2u}{dt^2} &\leq -2  &,\quad & \frac{d^2v}{dt^2} &\leq -2.
\end{alignat}

In all scenarios, the vehicles are also constrained to ensure a smooth trajectory and vehicle stability. To achieve this, according to the literature, the steering angle and steering rate need to be bounded~\cite{papaioannou2024reducing}:
\begin{align}
|\sigma| &\leq \frac{\pi}{9} \textup{rad}\\
\left|\frac{d\sigma}{dt}\right| &\leq \frac{\pi}{12} \textup{rads}^\inv.
\end{align}
A feature of the traction versus slip curve is that the traction initially increases as the slip increases, up to a certain point where the traction decreases as the slip is increases.
Hence, it is also important to control the wheel torque to ensure the vehicle does not slip beyond the point where traction starts decreasing. To understand why, we examine the relationship between input torque and wheel angular velocity, $d\omega/dt = \brac{T-Fr_e}J_w^\inv$. For a vehicle operating at the point where traction is maximized, if the input torque $T$ exceeds the maximum traction torque $F_{\max}r_e$, then $d\omega/dt$ will be positive, and the wheel's angular velocity will continuously increase. This leads to an increase in slip, causing a reduction in traction force, which further increases the wheel's angular acceleration. As a result, the slip continues to increase indefinitely, leading to a loss of control. Hence, to maintain stability and efficiency, we limit the slip to ensure the friction coefficient remains below $85\%$ of its maximum value. These bounds vary for hard and soft tires, as hard tires reach their peak friction coefficient at higher slip values.
\begin{alignat}{5}
     |\lambda_s| &\leq 0.034 &,\quad & |\alpha_s|   &\leq 0.041\\
     |\lambda_h|    &\leq 0.057 &,\quad &|\alpha_h| & \leq 0.073.
\end{alignat}
A limitation of the curvilinear model (see~\cref{sec:pre}) is that the vehicle must always have non-zero speed for $dt/ds$ to be finite. Hence, for the simulations, we limit the lower speed of the vehicle at $1\text{ms}^\inv$, in addition to a maximum speed limit $v_{\text{limit}}$, so $1\leq v \leq v_{\text{limit}}$.

The base vehicle is first simulated for each scenario. Next, the low-wear vehicle—equipped with a controller that optimizes for emissions while mimicking the base vehicle's performance—is simulated. The simulation generates a time series of the vehicle trajectory, including data such as emissions, lateral coordinates, torque input, and more. These outputs are then compared between the two vehicles.


\subsection{Simulation results}
\label{sec:simres}
\subsubsection{Emergency braking}
The first scenario, tests the vehicles in an emergency braking situation. Both vehicles repeat the simulation, each time starting at a different initial velocity $v_{\text{init}} \in \{30, 60, 120\}~\text{kmh}^\inv$, then attempt to come to a stop as quickly as possible. To simulate this, we solve the optimal control problem~\cref{eq:opti}, where the objective is to minimize the speed, i.e., $\cJ(\cQ,\cU) = \sum_{i=0}^d v(i)^2$. Note that the optimization over the path, here, acts as an automated method to generate the driver input and trajectory of the vehicles, given the all the driving and driver conditions described in the previous part.  The stopping distances and maximum decelerations are recorded in~\cref{tab:stop}.
\begin{table}[H]
\centering
\caption{Stopping distance and peak deceleration of test vehicles during emergency braking}
\label{tab:stop}
\begin{tabular}{cccccc}
\toprule
\multirow{2}{*}{\begin{tabular}[t]{@{}c@{}}Condition\\ Scaling \\ ($\zeta$)\end{tabular}} & \multirow{2}{*}{\begin{tabular}[t]{@{}c@{}}Initial\\ Velocity\\ (kmh$^{-1}$)\end{tabular}} & \multicolumn{2}{c}{\begin{tabular}[t]{@{}c@{}}Stopping Distance\\ (m)\end{tabular}} & \multicolumn{2}{c}{\begin{tabular}[t]{@{}c@{}}Peak Deceleration\\ (ms$^{-2})$\end{tabular}} \\
 &  & Base & Low-wear & Base & Low-wear \\
\midrule
\multirow{3}{*}{0.5} 
 & 30 & 6.3 & 7.0 & 5.4 & 4.9 \\
 & 60 & 25.8 & 28.6 & 5.5 & 5 \\
 & 120 & 98.6 & 109.6 & 5.9 & 5.4 \\
\midrule
\multirow{3}{*}{1.0}
& 30 & 3.2 & 3.6 & 10.8 & 9.8 \\
 & 60 & 13.2 & 14.4 & 10.9 & 9.9 \\
 & 120 & 50.6 & 56.2 & 11.3 & 10.3\\
\bottomrule
\end{tabular}
\end{table}
From the result above, we see that the predicted increase in stopping distance previously made is~\cref{sec:limit} is accurate. In all scenarios, the increase in stopping distance is approximately $10\%$. 

\subsubsection{Straight path}
\begin{figure}[htbp]
    \centering
    \captionsetup{width=1\textwidth}
    
    \subfloat[$v = 30~\text{kmh}^{-1},\,\zeta = 0.5$]{{\includegraphics[width = 0.9\textwidth]{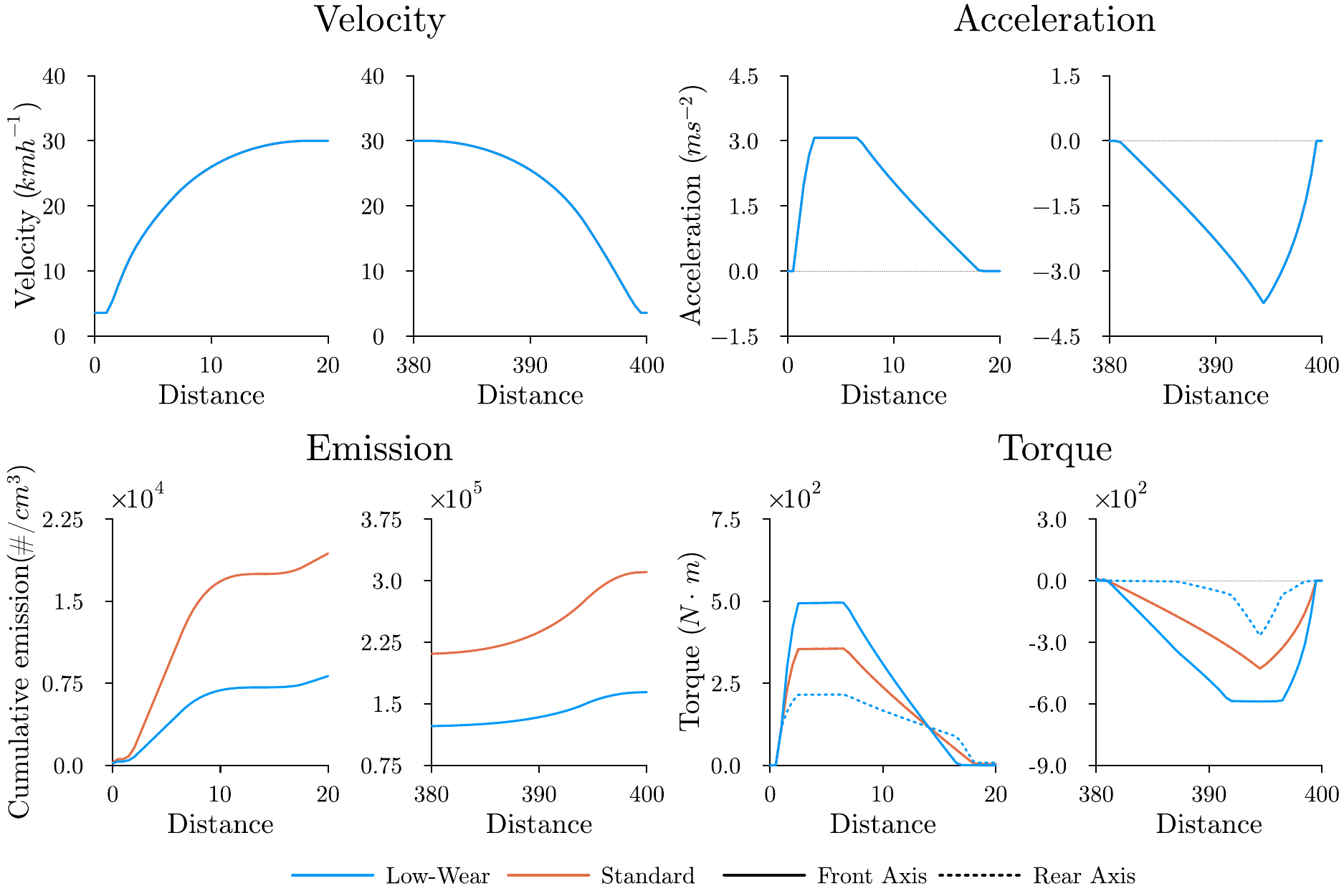} }}%
    \qquad
    \subfloat[$v = 30~\text{kmh}^{-1},\,\zeta = 1.0$]{{\includegraphics[width = 0.9\textwidth]{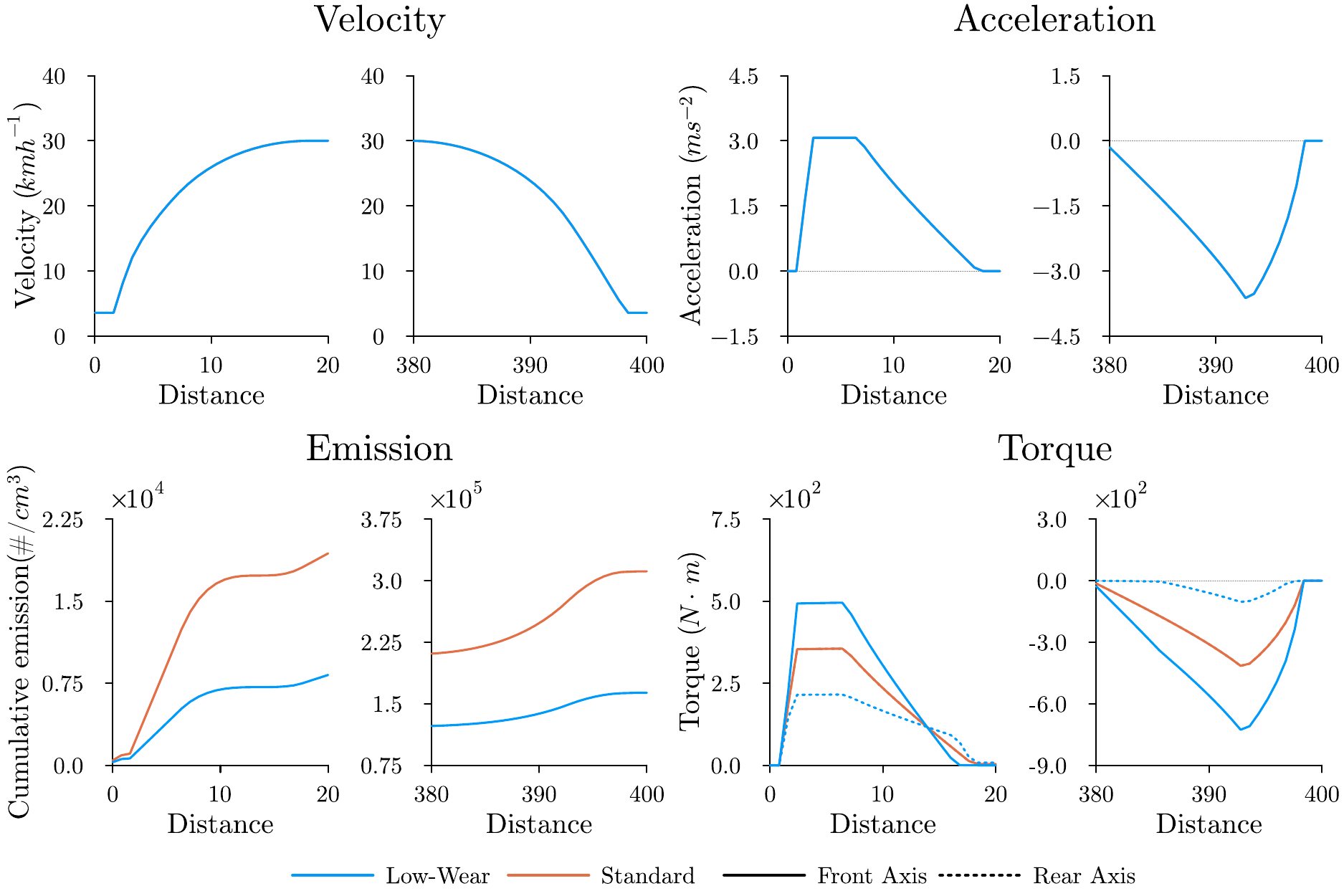} }}%
    \caption{\textbf{Straight trajectory simulation with the speed limit of 30 kmh$^\inv$}. The plots compare the velocity, acceleration, emission, and torque of the proposed low-wear vehicle to those of the base vehicle, simulated over a straight-line path with the speed limit at $30$ kmh$^\inv$. The top and bottom panels separate the results for (a) low-traction/wet ($\zeta=0.5$) and (b) normal traction ($\zeta=1$) conditions. The data is shown separately during acceleration and deceleration for better visualization.}
    \label{fig:straight-30}
\end{figure}

\begin{figure}[htbp]
    \centering
    \captionsetup{width=1\textwidth}
   
    \subfloat[$v = 60~\text{kmh}^{-1},\,\zeta = 0.5$]{{\includegraphics[width = 0.9\textwidth]{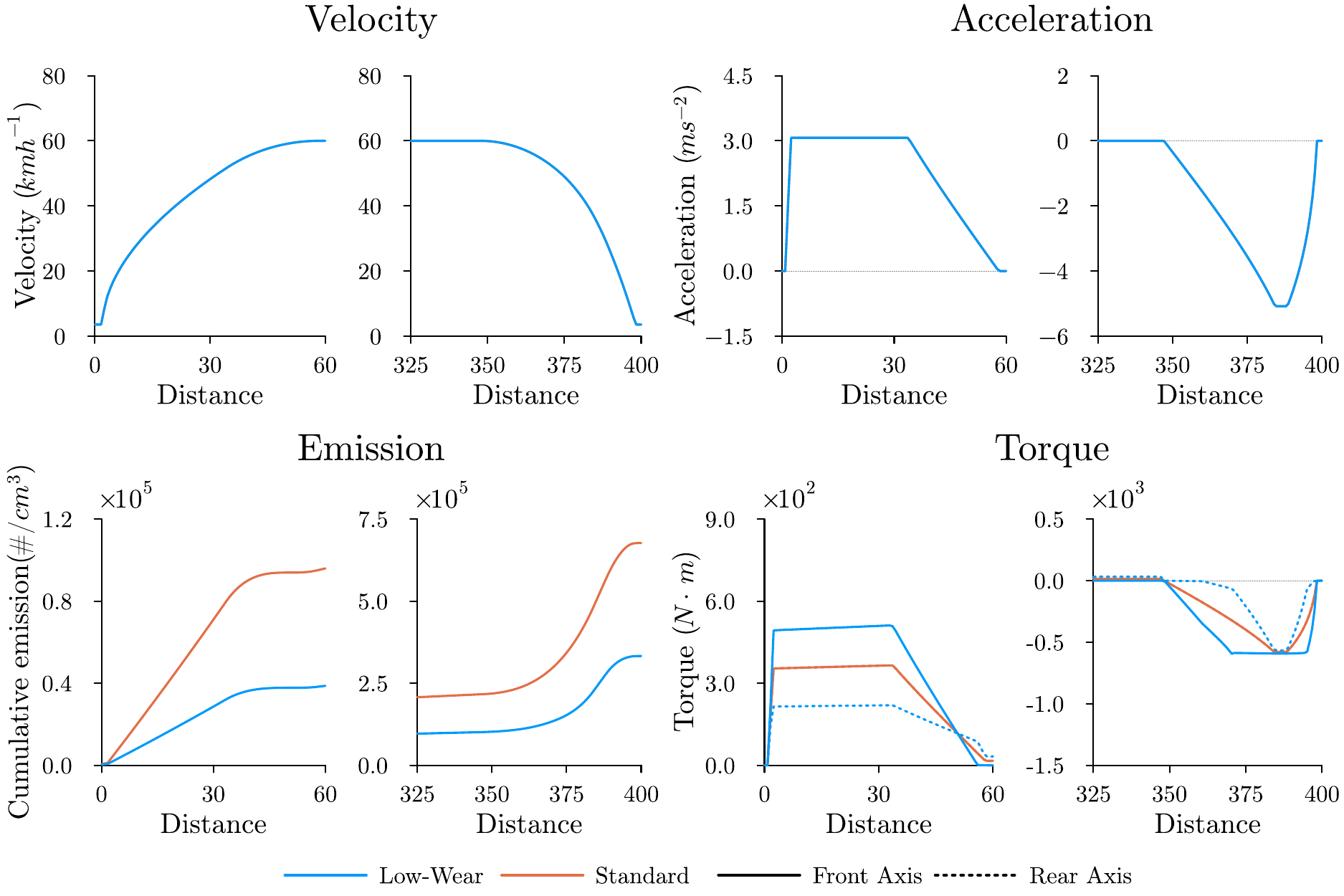} }}%
    \qquad
    \subfloat[$v = 60~\text{kmh}^{-1},\,\zeta = 1.0$]{{\includegraphics[width = 0.9\textwidth]{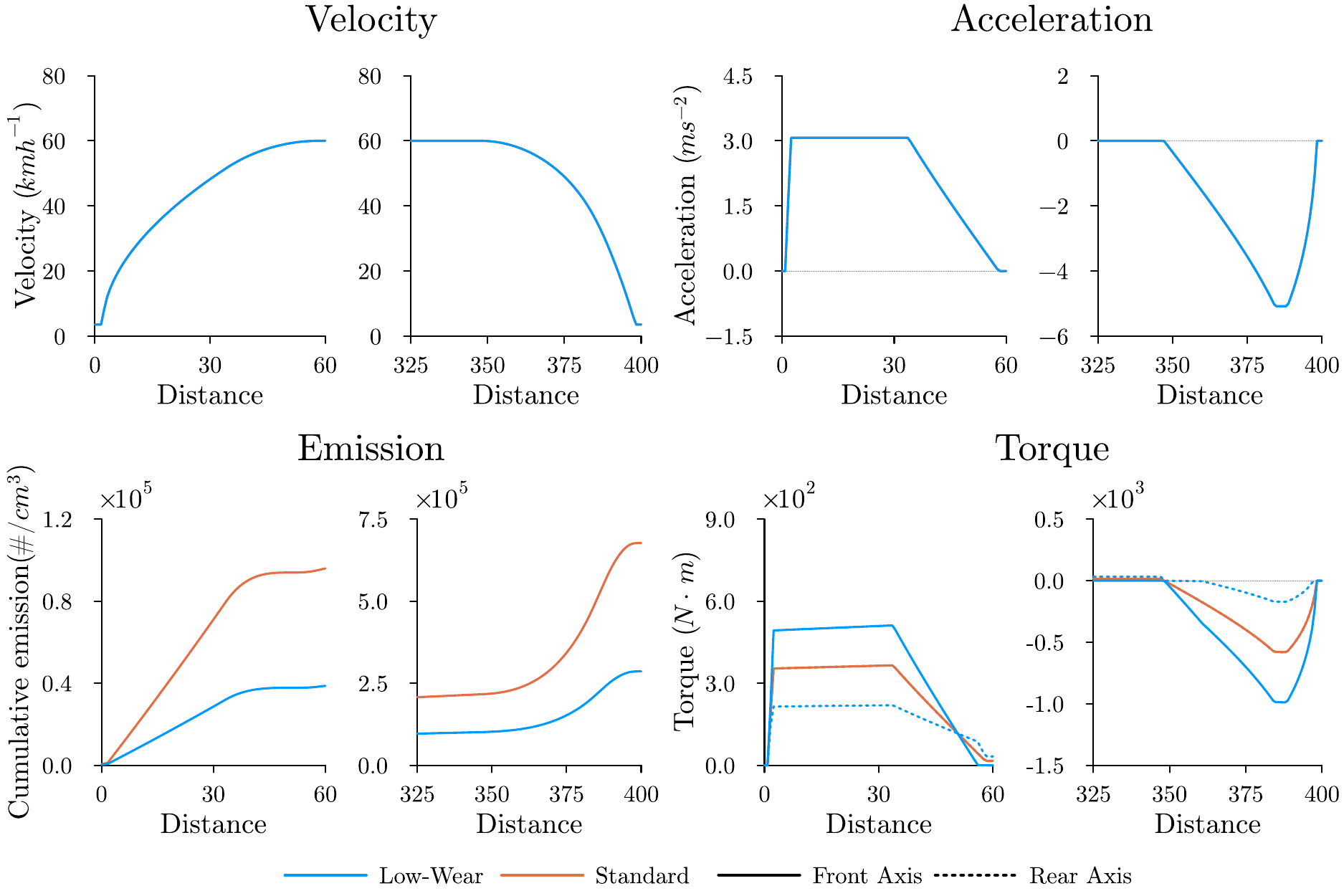} }}%
     \caption{\textbf{Straight trajectory simulation with the speed limit of 60 kmh$^\inv$}. The plots compare the velocity, acceleration, emission, and torque of the proposed low-wear vehicle to those of the base vehicle, simulated over a straight-line path with the speed limit at $60$ kmh$^\inv$. The top and bottom panels separate the results for (a) low-traction/wet ($\zeta=0.5$) and (b) normal traction ($\zeta=1$) conditions.}
        \label{fig:straight-60}
\end{figure}

\begin{figure}[htbp]
    \centering
    \captionsetup{width=1\textwidth}
   
    \subfloat[$v = 120~\text{kmh}^{-1},\,\zeta = 0.5$]{{\includegraphics[width = 0.9\textwidth]{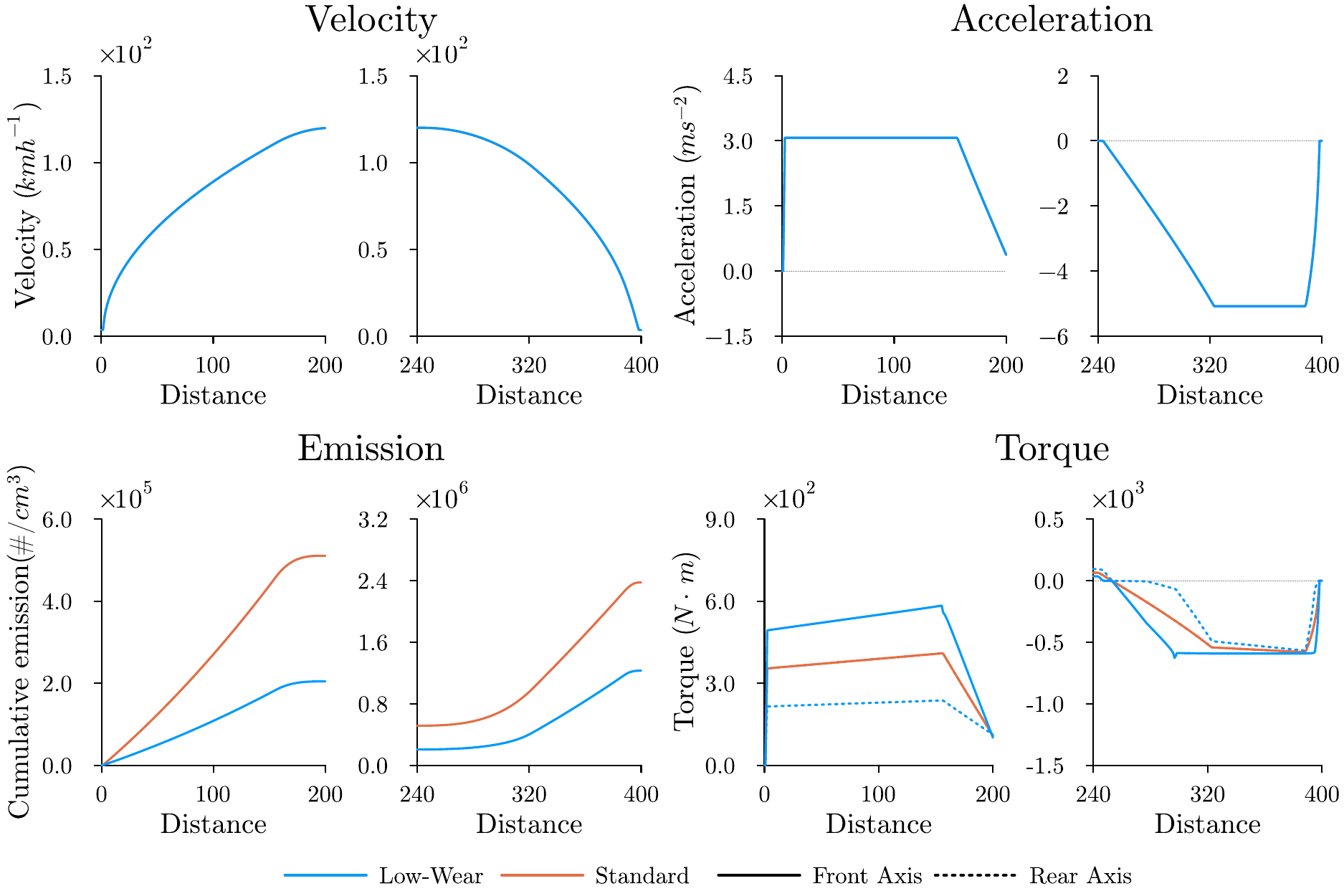} }}%
    \qquad
    \subfloat[$v = 120~\text{kmh}^{-1},\,\zeta = 1.0$]{{\includegraphics[width = 0.9\textwidth]{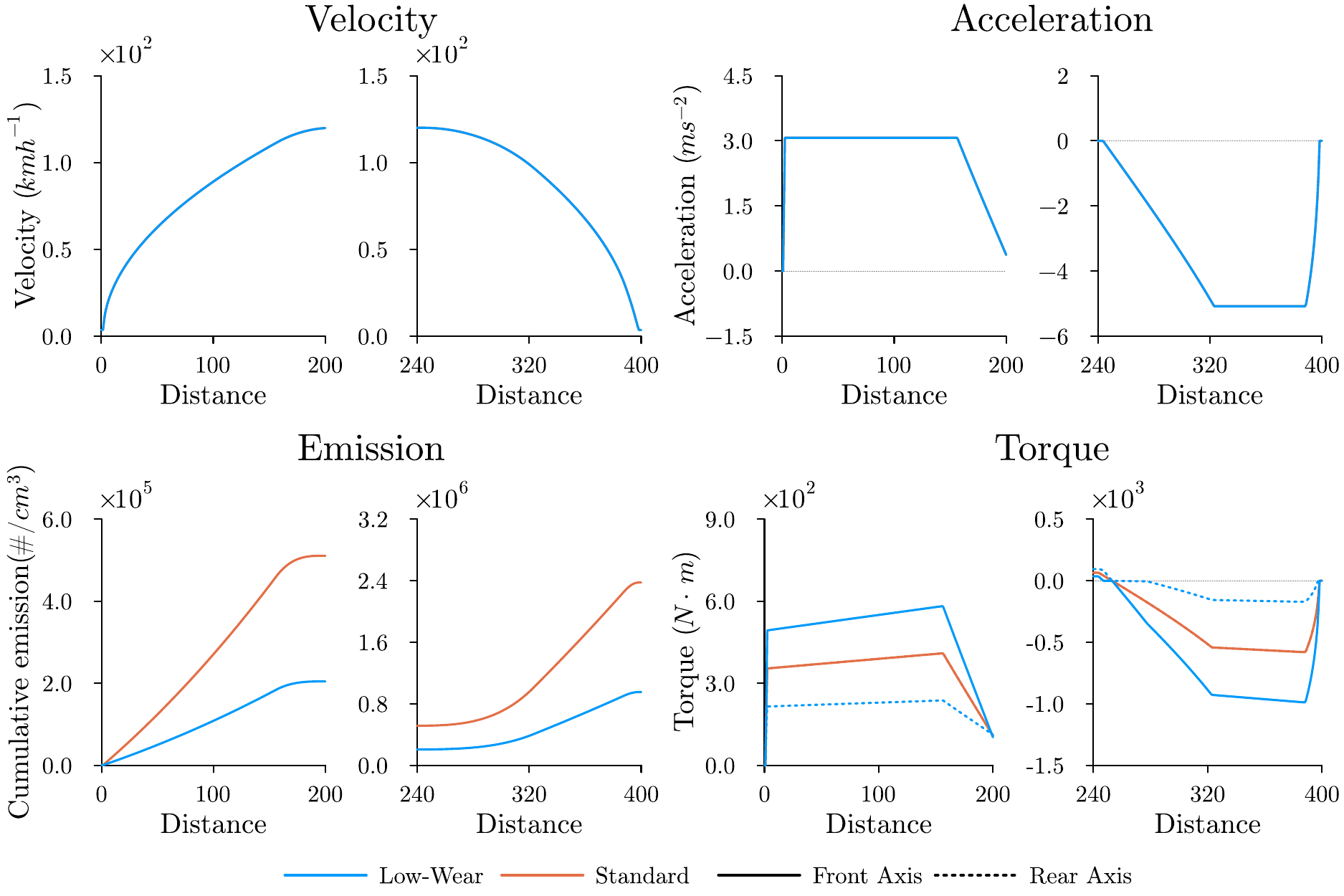} }}%
     \caption{\textbf{Straight trajectory simulation with the speed limit of 120 kmh$^\inv$}. The plots compare the velocity, acceleration, emission, and torque of the proposed low-wear vehicle to those of the base vehicle, simulated over a straight-line path with the speed limit at $120$ kmh$^\inv$. The top and bottom panels separate the results for (a) low-traction/wet ($\zeta=0.5$) and (b) normal traction ($\zeta=1$) conditions.}
        \label{fig:straight-120}
\end{figure}

In this scenario, we tested the vehicle on a straight path with length of $400$ m, where each vehicle will accelerate to $30~\text{kmh}^\inv$, $60~\text{kmh}^\inv$, and $120~\text{kmh}^\inv$ in separate runs of the simulation and then decelerate to a stop as quickly as possible. Conditions stated in \ref{sec:setting}, such as staying in the comfortable limit from Bae et al.~\cite{bae2019toward,bae2020self} are constraining the simulations at all times. 
The main objective is to evaluate the vehicle's traction and braking performance while also demonstrating the emission-saving capability of the proposed system during periods of high tire emissions caused by elevated forces.
Simulation results are presented in~\cref{fig:straight-30},~\cref{fig:straight-60},and ~\cref{fig:straight-120}, corresponding to scenarios with maximum speeds of $30,60$ and $120$ kmh$^\inv$, respectively. Each figure includes two panels,(a) and (b), representing the two driving condition scenarios. Within each panel, comparisons are shown for four parameters: velocity, acceleration, emissions, and torque.


In all of these figures, we use orange to denote the vehicle with the standard tire setup and blue to denote our low-wear setup. In the emission and torque subplots, the data is shown separately for the front axis (solid line) and the rear axis (dotted line). The tire particle emissions calculated via the simulations of the straight-path scenario are summarized in~\cref{tab:emit-straight}.

For better visualization of the results in each subplot of~\cref{fig:straight-30,fig:straight-60,fig:straight-120}, the velocity, acceleration, emission, and torque are separately shown during acceleration and deceleration.
The results show that the proposed dual tire-profile vehicle equipped with the control system can match the driving performance of the vehicle with all high-performing tires in all conditions. (In all \enquote{velocity} and \enquote{acceleration} plots, the blue curves lie exactly over the orange curves.) To explain this, it is useful to consider that the performance needs to surpass the driver's comfort limits before reaching the tire's performance limit. Below these limits, however, our control method is capable of tuning the performance of the low-wear vehicle and matching that of the base vehicle. This means that the driver will notice no difference in the driving experience in the low-wear vehicle compared to the base vehicle.

Regarding emission, the proposed vehicle can reduce the tire particle generation by at least $48 \%$ in all conditions at all times (compare the red and blue curves in \enquote{Emission} subplots of Figs.~\ref{fig:straight-30}, \ref{fig:straight-60}, and \ref{fig:straight-120}). As it can be seen in \enquote{torque} subplots, the controller consistently divides the torque unevenly on the front and rear axes of the low-wear vehicle. Optimally dividing the torque (with a smaller magnitude on the rear axis) is able to successfully reduce the total TWP emissions and simultaneously tune the low-wear vehicle's performance as if all tires are soft high-performance tires. In normal traction condition where $\zeta = 1.0$ (panel (b) of the figures), the low-wear vehicle saves up to $60\%$ in emissions, and the savings grow as the speed increases. However, we do not observe this increase in emission-saving performance with the speed in low-traction conditions where $\zeta = 0.5$ (panel (a) of the figures). This is because the hard tire reaches the traction force limits earlier in the low-traction case, and the dual tire-profile (low-wear) vehicle must rely more on the soft tires.

\begin{table}[htbp]
\centering
\caption{Emission comparison of test vehicles on the straight path.}
\label{tab:emit-straight}
\begin{tabular}{ccccc}
\toprule
\multirow{2}{*}{\begin{tabular}[t]{@{}c@{}}Condition\\ Scaling \\ ($\zeta$)\end{tabular}} & \multirow{2}{*}{\begin{tabular}[t]{@{}c@{}}Initial\\ Velocity\\ (kmh$^{-1}$)\end{tabular}} & \multicolumn{2}{c}{\begin{tabular}[t]{@{}c@{}}Particle Number\\ (\#/cm$^3$)\end{tabular}} & \multirow{2}{*}{\begin{tabular}[t]{@{}c@{}}Emission \\  Reduction\\ (\%)\end{tabular}} \\
 &  & Base & Low-wear & \\
\midrule
\multirow{3}{*}{0.5} 
 & 30 & $3.1\cdot 10^5$ & $1.6\cdot 10^5$ & 48  \\
 & 60 & $6.8\cdot 10^5$ & $3.3\cdot 10^5$ & 51 \\
 & 120 & $2.4\cdot 10^6$ & $1.2\cdot 10^6$& 48  \\
\midrule
\multirow{3}{*}{1.0}
& 30 & $3.1\cdot 10^5$ & $1.6\cdot 10^5$ &48 \\
 & 60 & $6.7\cdot 10^5$ & $2.9\cdot 10^5$ & 57  \\
 & 120 & $2.4\cdot 10^6$ & $9.5\cdot 10^5$ & 61 \\
\bottomrule
\end{tabular}
\end{table}

\subsection{Curved path}

In the last scenario, the vehicles are simulated on curved paths.  
This scenario is important for testing the vehicle's handling ability, at different speeds with significant lateral speed. In this set of simulations, the vehicle first travels a straight section of $100$ m, then follows a circular arc, and finishes the path by traveling on another $100$ m straight section. For each vehicle type, namely, the base and the low-wear vehicle, we repeat three runs of the simulation, corresponding to different constant speeds and radii of the circular section. The speed-radius pairs selected for different runs of the simulation are: i) $30\,\text{kmh}^\inv$ speed and $32$ m radius, ii) $60\,\text{kmh}^\inv$ speed and $127$ m radius, and iii) $120\,\text{kmh}^\inv$ speed and $510$ m radius. These selected curvatures are two steps below the desirable minimum radius for each speed, based on \emph{The Design Manual for Roads and Bridges} published by National Highways of the United Kingdom; see Table 2.10, CD 109 in~\cite{dmrb}. In other words, the radii in our scenarios are half the standard desirable minimum radii for the selected speed, allowing to test the performance of the vehicles, especially the proposed low-wear vehicle under extreme driving conditions.

\begin{table}[b]
\centering
\caption{Emission comparison of test vehicles in curved trajectory}
\label{tab:emit-curve}
\begin{tabular}{ccccc}
\toprule
\multirow{2}{*}{\begin{tabular}[t]{@{}c@{}}Condition\\ Scaling \\ ($\zeta$)\end{tabular}} & \multirow{2}{*}{\begin{tabular}[t]{@{}c@{}}Initial\\ Velocity\\ (kmh$^{-1}$)\end{tabular}} & \multicolumn{2}{c}{\begin{tabular}[t]{@{}c@{}}Particle Number\\ (\#/cm$^3$)\end{tabular}} & \multirow{2}{*}{\begin{tabular}[t]{@{}c@{}}Emission \\  Reduction\\ (\%)\end{tabular}} \\
 &  & Base & Low-wear & \\
\midrule
\multirow{3}{*}{0.5} 
 & 30 & $2.1\cdot 10^5$ & $1.1\cdot 10^5$ & 46  \\
 & 60 & $3.9\cdot 10^5$ & $2.0\cdot 10^5$ & 50 \\
 & 120 & $3.1\cdot 10^5$ & $1.1\cdot 10^5$& 63  \\
\midrule
\multirow{3}{*}{1.0}
& 30 & $2.1\cdot 10^5$ & $1.1\cdot 10^5$ & 46 \\
 & 60 & $4.1\cdot 10^5$ & $2.0\cdot 10^5$ & 50  \\
 & 120 & $3.4\cdot 10^5$ & $1.3\cdot 10^5$ & 61 \\
\bottomrule
\end{tabular}
\end{table}

\begin{figure}[H]
    \centering
    \captionsetup{width=0.8\textwidth}
    \subfloat[$v = 30~\text{kmh}^{-1},\,\zeta = 0.5$]{{\includegraphics[width = \textwidth]{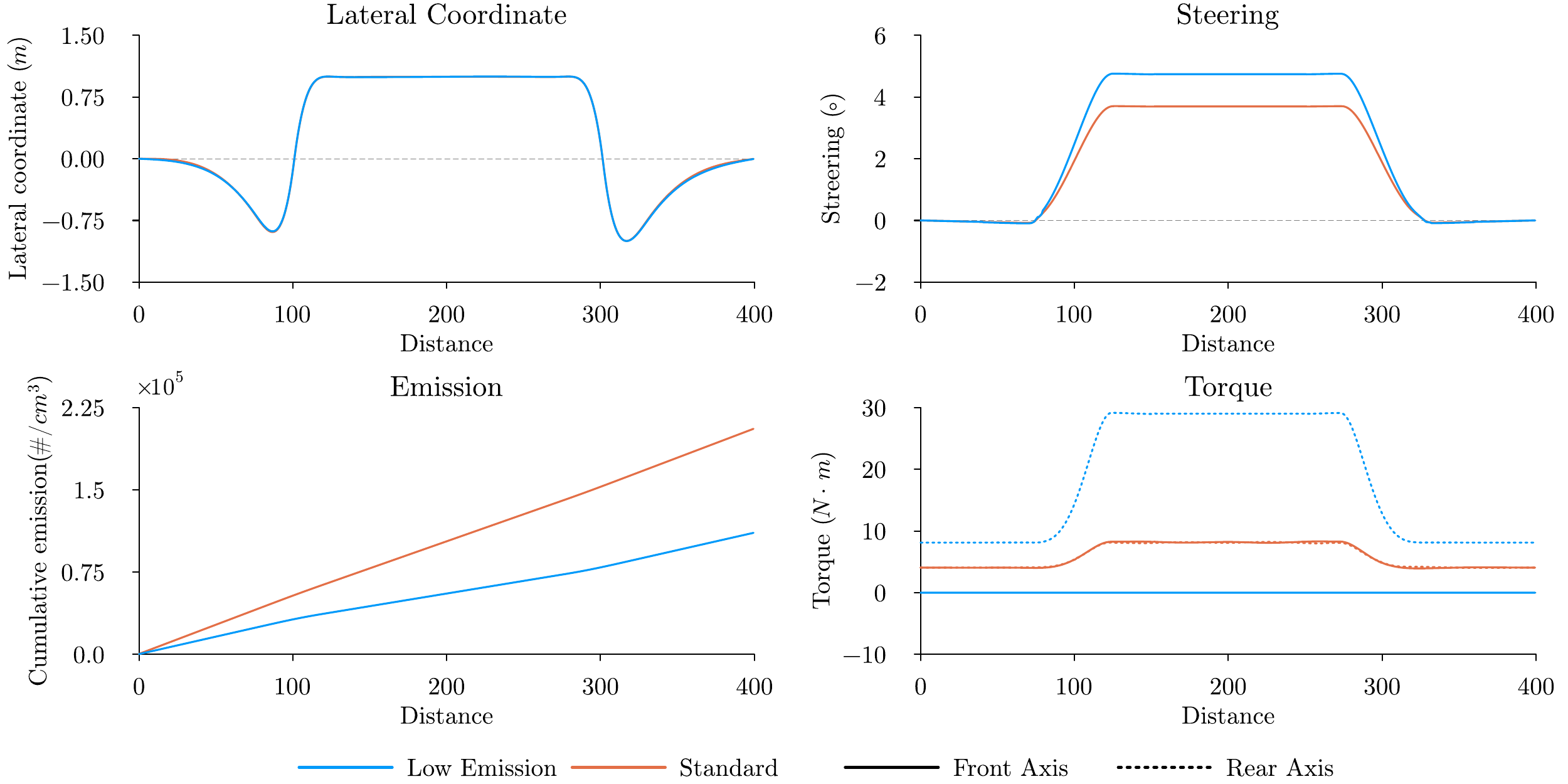} }}%
    \qquad
    \subfloat[$v = 30~\text{kmh}^{-1},\,\zeta = 1.0$]{{\includegraphics[width = \textwidth]{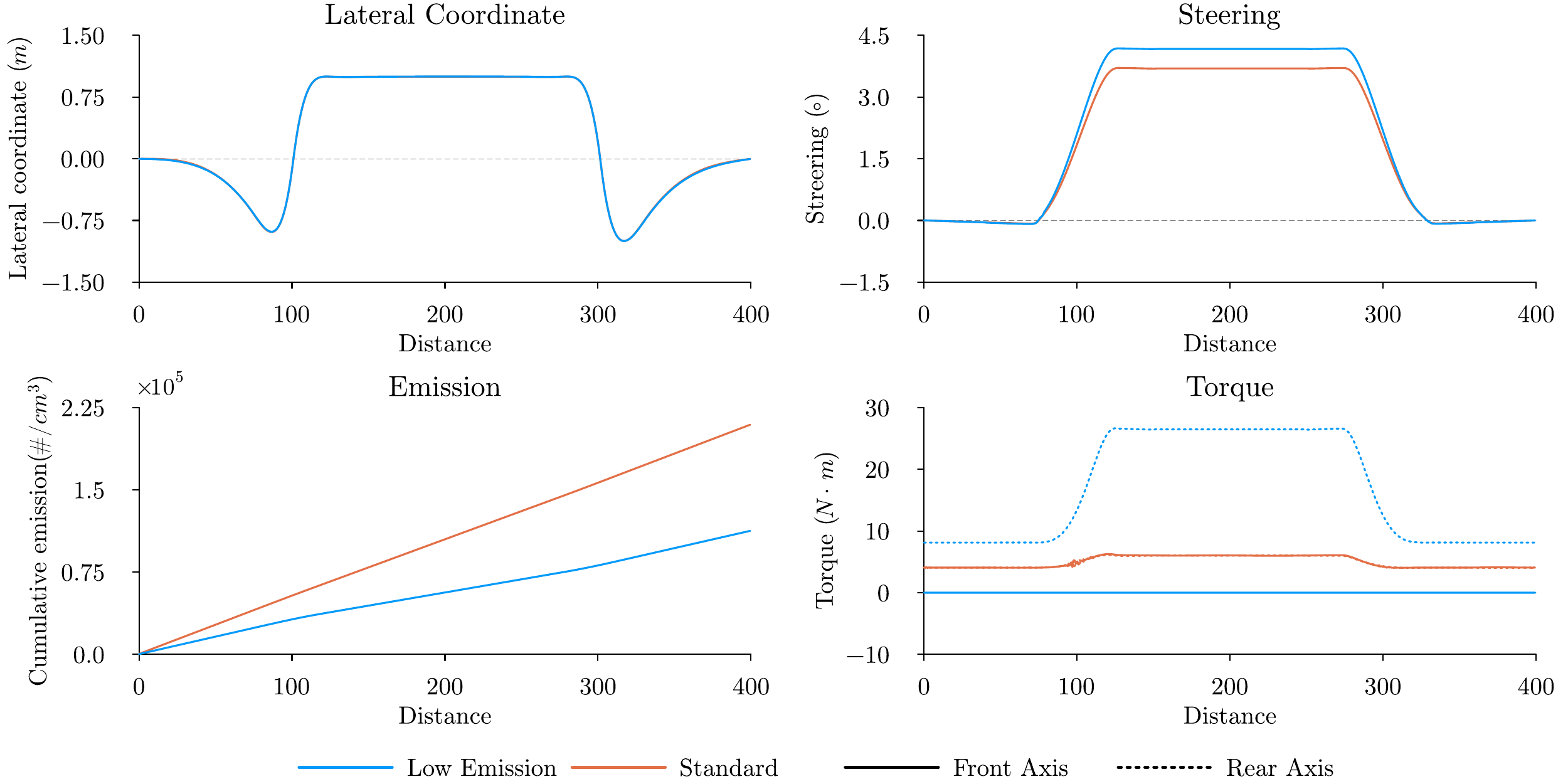} }}%
    
    \caption{\textbf{Curved trajectory simulation with the speed limit at 30 kmh$^{-1}$}.  The plots compare the lateral coordinate, steering angle, emission, and torque of the proposed low-wear vehicle to those of the base vehicle, simulated over a straight-line path with the speed limit at $60$ kmh$^\inv$. The top and bottom panels separate the results for (a) low-traction/wet ($\zeta=0.5$) and (b) normal traction ($\zeta=1$) conditions.}
        \label{fig:curve-30}
\end{figure}

\begin{figure}[H]
    \centering
    \captionsetup{width=0.8\textwidth}
    \subfloat[$v = 60~\text{kmh}^{-1},\,\zeta = 0.5$]{{\includegraphics[width = \textwidth]{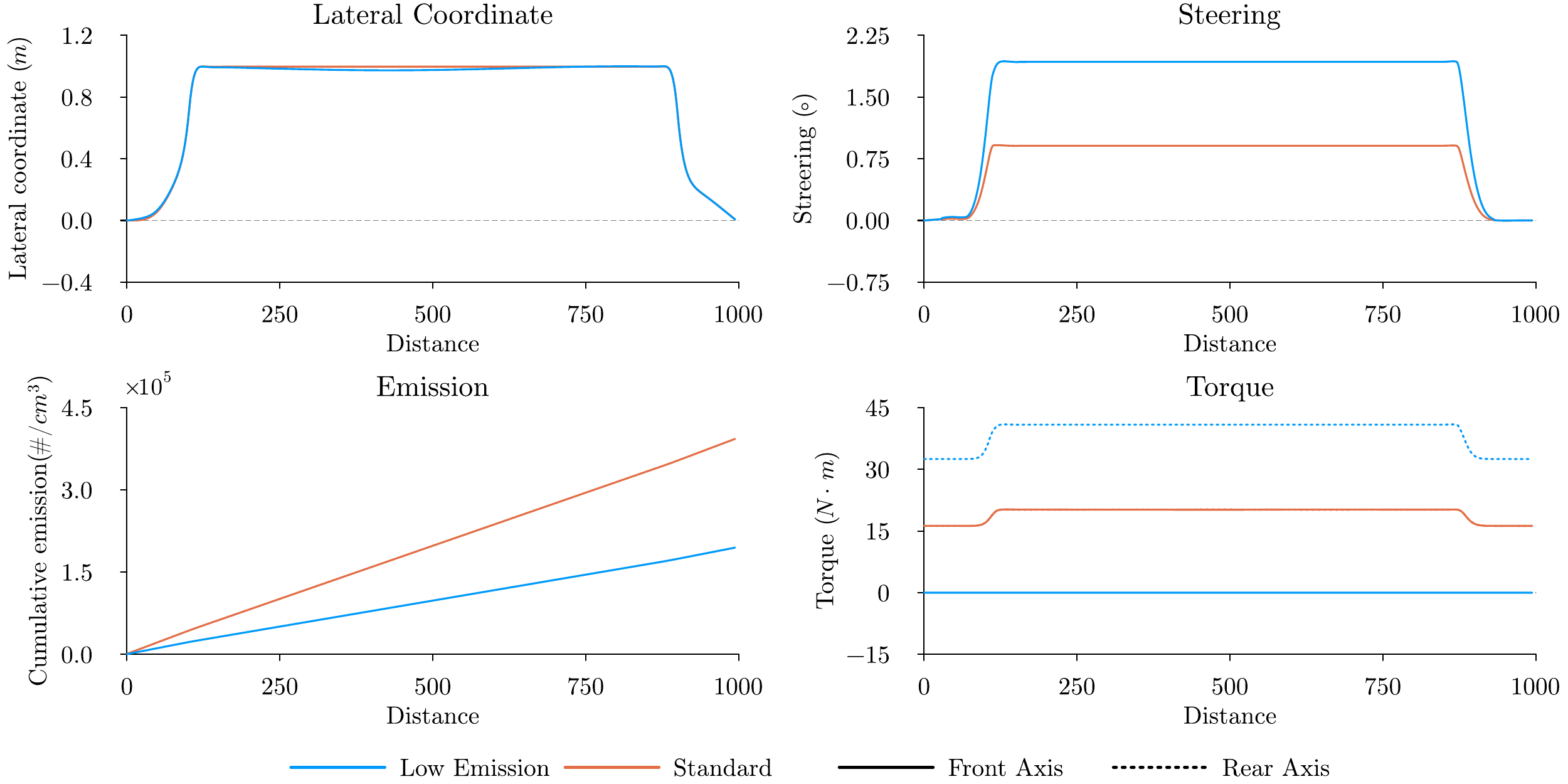} }}%
    \qquad
    \subfloat[$v = 60~\text{kmh}^{-1},\,\zeta = 1.0$]{{\includegraphics[width = \textwidth]{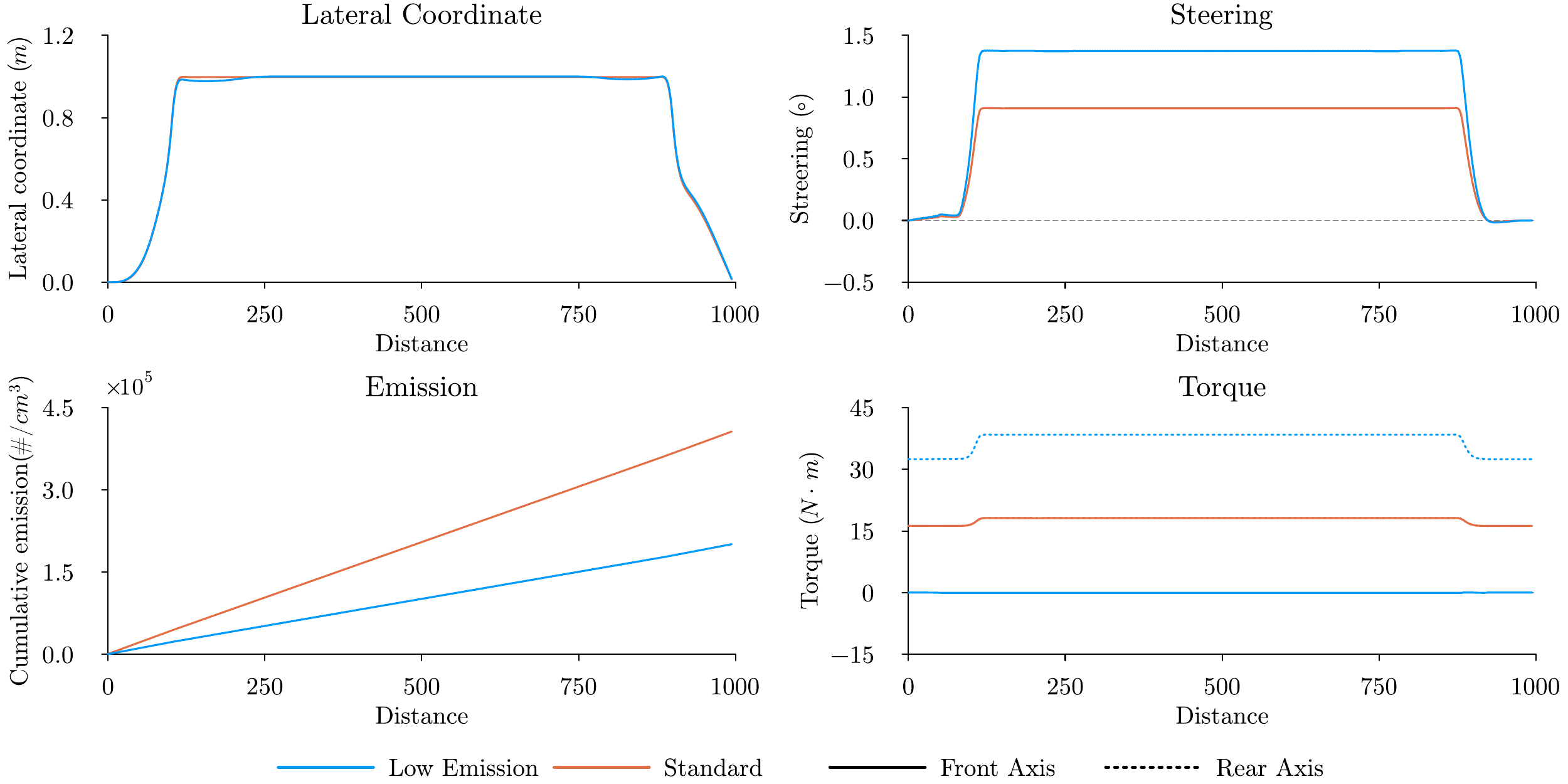} }}%
    \caption{\textbf{Curved trajectory simulation with the speed limit at 60 kmh$^{-1}$}. The plots compare the lateral coordinate, steering angle, emission, and torque of the proposed low-wear vehicle to those of the base vehicle, simulated over a straight-line path with the speed limit at $60$ kmh$^\inv$. The top and bottom panels separate the results for (a) low-traction/wet ($\zeta=0.5$) and (b) normal traction ($\zeta=1$) conditions.}
        \label{fig:curve-60}
\end{figure}

\begin{figure}[H]
    \centering
    \captionsetup{width=0.8\textwidth}
    \subfloat[$v = 120~\text{kmh}^{-1},\,\zeta = 0.5$]{{\includegraphics[width = \textwidth]{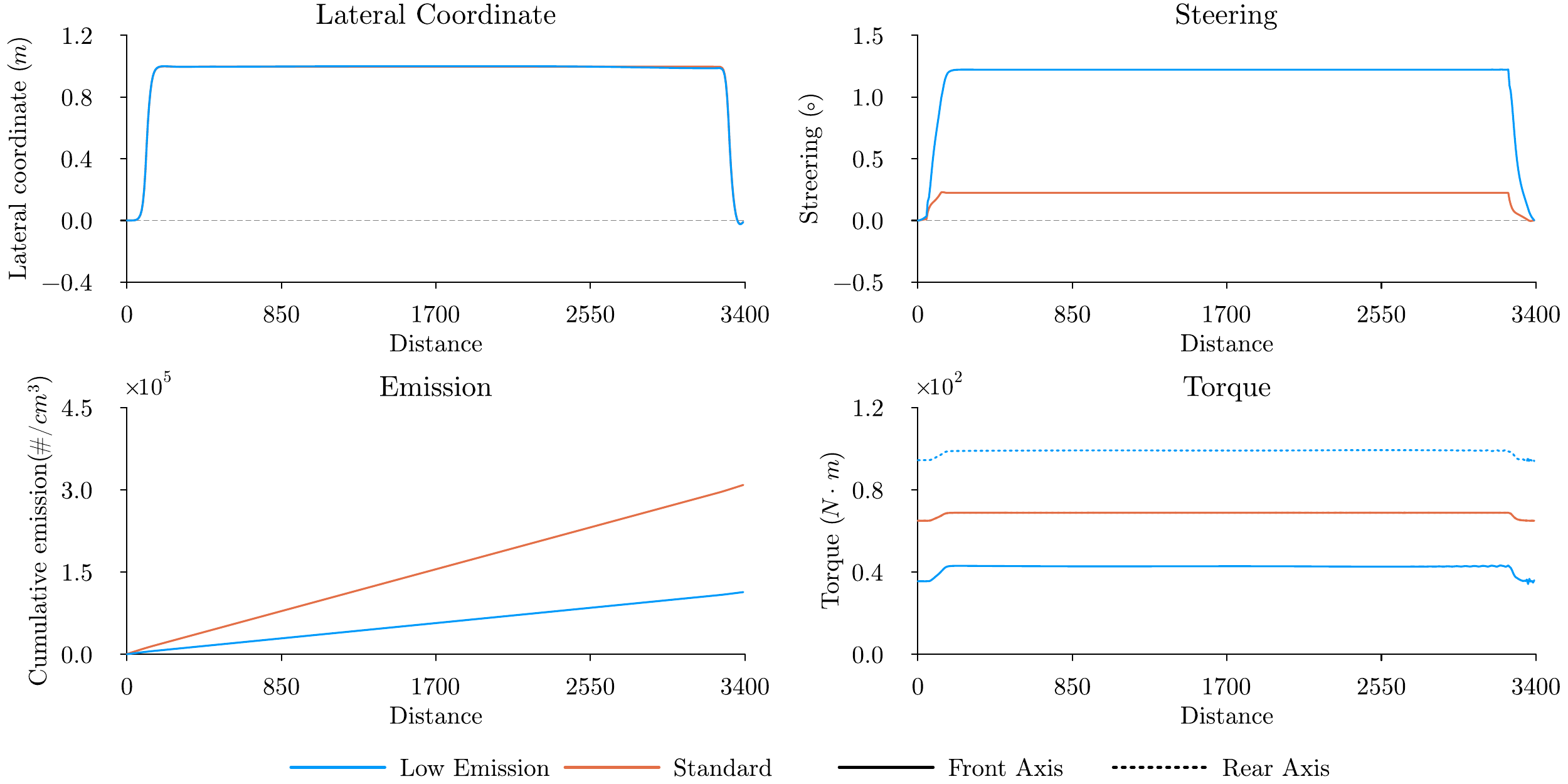} }}%
    \qquad
    \subfloat[$v = 120~\text{kmh}^{-1},\,\zeta = 1.0$]{{\includegraphics[width = \textwidth]{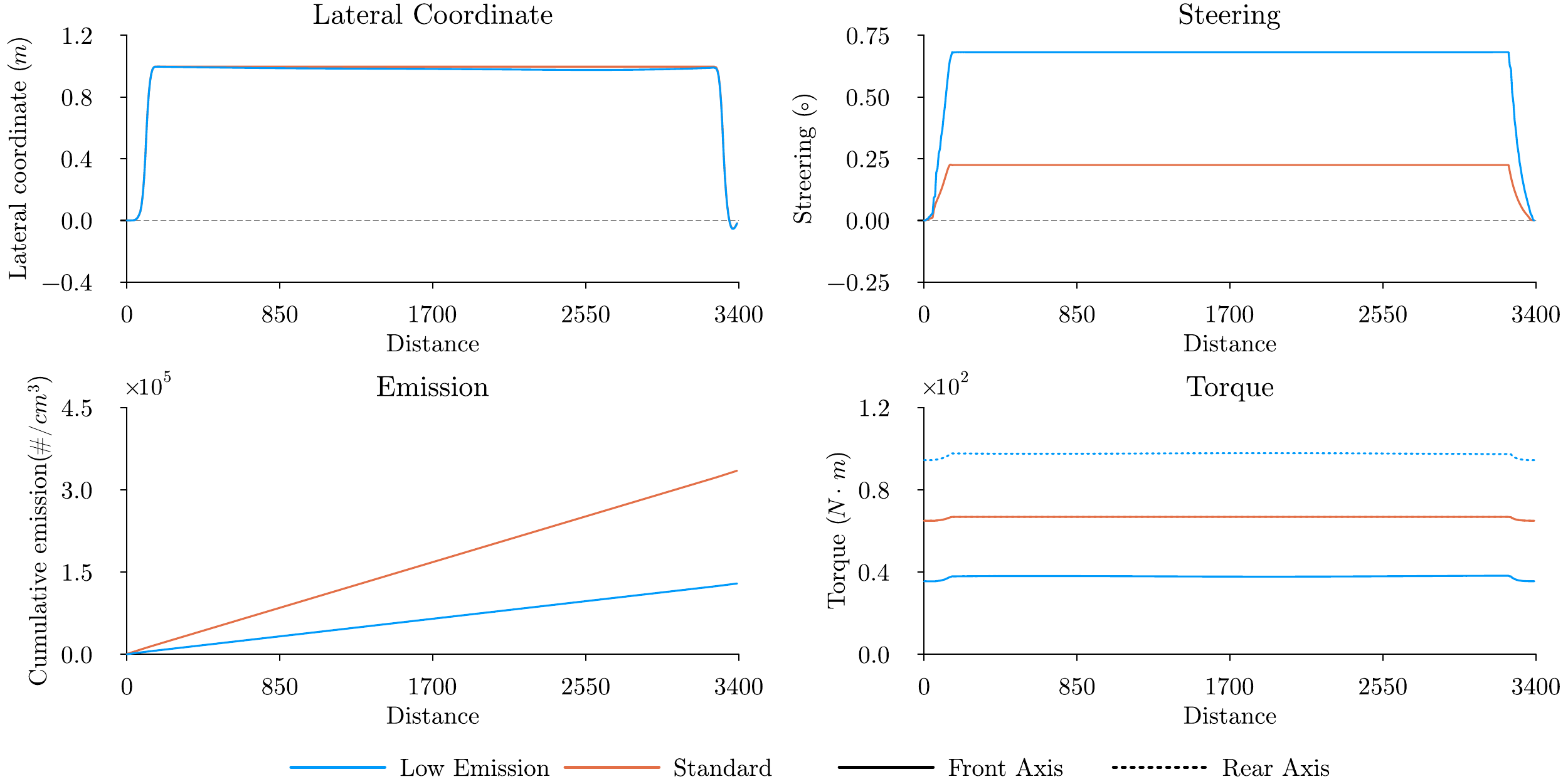} }}%
     \caption{\textbf{Curved trajectory simulation with the speed limit at 120 kmh$^{-1}$}. The plots compare the lateral coordinate, steering angle, emission, and torque of the proposed low-wear vehicle to those of the base vehicle, simulated over a straight-line path with the speed limit at $120$ kmh$^\inv$. The top and bottom panels separate the results for (a) low-traction/wet ($\zeta=0.5$) and (b) normal traction ($\zeta=1$) conditions.}
        \label{fig:curve-120}
\end{figure}

The simulation results are presented in Figs~\ref{fig:curve-30}, \ref{fig:curve-60}, and \ref{fig:curve-120} separately for different speed and curve radius pairs. Each figure is divided into panels (a) low-traction (or wet) condition $\zeta=0.5$, and (b) normal traction condition $\zeta=1$, where we present the lateral force, steering, emission, and torque values of the vehicles over the distance along the path. In all visualizations, orange to color-codes the base vehicle with the standard tire set-up and blue indicates our proposed low-wear controlled vehicle with the dual tire-profile set-up. In the torque subplots, we use solid and dotted lines to show the output from the front and rear axes, respectively. 

The results show that controlling the steering and torque division can successfully lead to the low-wear vehicle replicating the trajectory of the base vehicle despite the former using hard tires on one axis. This is evident in the insignificant difference in the lateral coordinates of the two vehicles over the whole path in all of the results. Also, see the steering and torque subplots depicting the role of the control system in compensating for the lower performance of the hard tires in the lower-wear vehicle. Note that the drivers' steering inputs over the path are equal between the two vehicles, and the controller is responsible for slight adjustment of the angle in the low-wear vehicle.

In the curved path scenario, we observe a division of force that may seem counterintuitive at first. In the low-wear vehicle, the controller puts more force on the soft tires on the rear axis. This is due to the data-driven realistic tire wear model used here. It can be seen in~\cref{fig:emissionfit} that the optimal torque to minimize emission for the model is not zero but rather approximately at $500$ N. Hence, to minimize the emission, the force on the soft tire has to be close to $500$ N rather than $0$. As a result, when the total force is small, to minimize the emissions (more easily generated by the soft tire), the controller correctly allocates more torque to the rear axis, which puts the force on the soft tire to be closer to the emission-optimal value. This is not observed in the straight path scenario since the 

Regarding emission, the proposed low-wear vehicle equipped with the control system saves between $45\%$ to $60\%$ in total tire-wear emissions of the base vehicle (with all soft tires); the emission reduction is more at higher speeds. The total tire wear particle emissions calculated in the simulations of this scenario are summarized in~\cref{tab:emit-curve}.


\section{Conclusion}
\label{sec:con}
In this work, we presented a simple yet effective approach to reducing tire emissions in all-wheel-drive electric vehicles. By equipping one wheel axis with a low-wear, low-traction tire, and utilizing the appropriate control in the vehicle, we achieved a significant reduction in tire emissions. The proposed model was tested across several scenarios and different conditions to demonstrate the effectiveness of the proposed system. Our control system unevenly distributes the torque and thus tire force to increase the utilization of the low-wear tire when possible. Additionally, based on a mathematical model of vehicle dynamics, we demonstrated that the vehicle with dual tire profiles can be controlled in a way that preserves handling characteristics similar to a standard vehicle with identical tire profiles in most driving scenarios and even in extreme driving conditions. 
Our control algorithm is both simple and effective, as demonstrated by numerical experiments showing a significant reduction in TWP emissions while maintaining vehicle stability and performance. Given the serious environmental and health risks posed by TWP—including air and water pollution and the release of harmful microplastics—this work supports a critical research direction aimed at reducing environmental impact, protecting public health, and advancing sustainable transportation.

An important feature of our algorithm is that it is not sensitive to the choice of tire wear model, and it will be applicable given any model of tire wear as a function of the longitudinal force. This is practically useful as consensus has not been reached in previous works regarding a universal tire wear model, and a variety of tire wear models have been used in the literature~\cite{foitzik2018investigation,gao2022torque,obereigner2020low,papaioannou2022optimal}.

Yet, a limitation of our approach is that the control algorithm requires sufficient accuracy in the tire model for computation. However, these measurements are not always readily available and may also change in tires over time as the tires wear. Machine learning control is a suitable approach to address this challenge. In particular, reinforcement learning offers a promising solution for controlling complex dynamic systems while adapting to changes. The primary drawback of machine learning methods, however, is their reliance on trial-and-error learning, which is unsuitable for safety-critical control systems where errors are not tolerated~\cite{henderson2018deep,recht2019tour,cheng2019control}. Recently, there have been many advances in safe reinforcement learning~\cite{tianreinforcement,gu2022constrained,isele2018safe}, where maintaining the safety of the system is essential and cannot be compromised during the learning process. Leveraging the most recent advances in machine learning in designing adaptive control systems that guarantee vehicles' stability and performance, along with optimizing the emissions, is an interesting direction for future research with huge benefits.

\def\bibsection{} 
\bibliographystyle{ieeetr}  
\bibliography{references} 
\end{document}